\documentclass{pasj01}
\usepackage[T1]{fontenc}

\Received{$\langle$reception date$\rangle$}
\Accepted{$\langle$acception date$\rangle$}
\Published{$\langle$publication date$\rangle$}

\begin{document}

\title{Clustering of quasars in a wide luminosity range at redshift 4 with Subaru Hyper Suprime-Cam wide field imaging}
\author{%
       Wanqiu \textsc{He}\altaffilmark{1}, 
       Masayuki \textsc{Akiyama}\altaffilmark{1},
       James \textsc{Bosch}\altaffilmark{2},
       Motohiro \textsc{Enoki}\altaffilmark{3},
       Yuichi  \textsc{Harikane}\altaffilmark{4,}\altaffilmark{5},
       Hiroyuki  \textsc{Ikeda}\altaffilmark{6},
       Nobunari \textsc{Kashikawa}\altaffilmark{6,}\altaffilmark{7},
       Toshihiro \textsc{Kawaguchi}\altaffilmark{8},
       Yutaka \textsc{Komiyama}\altaffilmark{6,}\altaffilmark{7},
       Chien-Hsiu  \textsc{Lee}\altaffilmark{9},
       Yoshiki \textsc{Matsuoka}\altaffilmark{10,}\altaffilmark{6},
       Satoshi \textsc{Miyazaki}\altaffilmark{6,}\altaffilmark{7},
       Tohru  \textsc{Nagao}\altaffilmark{10}, 
       Masahiro \textsc{Nagashima}\altaffilmark{11},
       Mana \textsc{Niida}\altaffilmark{12},
       Atsushi J \textsc{Nishizawa}\altaffilmark{13},
       Masamune   \textsc{Oguri}\altaffilmark{14,}\altaffilmark{5,}\altaffilmark{15},
       Masafusa \textsc{Onoue}\altaffilmark{6,}\altaffilmark{7},
       Taira \textsc{Oogi}\altaffilmark{15},
       Masami   \textsc{Ouchi}\altaffilmark{4},
       Andreas  \textsc{Schulze}\altaffilmark{6}, 
       Yuji \textsc{Shirasaki}\altaffilmark{6},
       John D. \textsc{Silverman}\altaffilmark{15},
       Manobu M. \textsc{Tanaka}\altaffilmark{16},
       Masayuki \textsc{Tanaka}\altaffilmark{6},
       Yoshiki \textsc{Toba}\altaffilmark{17},
       Hisakazu \textsc{Uchiyama}\altaffilmark{7},
       Takuji \textsc{Yamashita}\altaffilmark{10}}
   
 \altaffiltext{1}{Astronomical Institute, Tohoku University, Aramaki, Aoba, Sendai 980-8578, Japan }
 \altaffiltext{2}{Dept. of Astrophysical Sciences, Princeton University, 4 Ivy Lane, Princeton, NJ 08544, USA}
 \altaffiltext{3}{Faculty of Business Administration, Tokyo Keizai University, Kokubunji, Tokyo, 185-8502, Japan}
 \altaffiltext{4}{Institute for Cosmic Ray Research, The University of Tokyo, 5-1-5
Kashiwanoha, Kashiwa, Chiba 277-8582, Japan}
\altaffiltext{5}{Department of Physics, Graduate School of Science, The University
of Tokyo, 7-3-1 Hongo, Bunkyo, Tokyo, 113-0033, Japan}
\altaffiltext{6}{National Astronomical Observatory of Japan, Mitaka, Tokyo 181-8588, Japan}
\altaffiltext{7}{Department of Astronomy, School of Science, SOKENDAI (The Graduate University for
Advanced Studies), Mitaka, Tokyo 181-8588, Japan}
\altaffiltext{8}{Department of Economics, Management and Information Science, Onomichi City University, Hisayamada 1600-2, Onomichi, Hiroshima 722-8506, Japan}
\altaffiltext{9}{Subaru Telescope, NAOJ, 650 N Aohoku Pl, Hilo, HI 96720, USA}
\altaffiltext{10}{Research Center for Space and Cosmic Evolution, Ehime University, Matsuyama, Ehime 790-8577, Japan}
\altaffiltext{11}{Faculty of Education, Bunkyo University, Koshigaya, Saitama 343-8511, Japan}
\altaffiltext{12}{Graduate School of Science and Engineering, Ehime University, Bunkyo-cho 2-5, Matsuyama 790-8577, Japan}
\altaffiltext{13}{Institute for Advanced Research, Nagoya University, Chikusaku, Nagoya 464-8602, Japan}
\altaffiltext{14}{Research Center for the Early Universe, University of Tokyo, Tokyo
113-0033, Japan}
\altaffiltext{15}{Kavli Institute for the Physics and Mathematics of the Universe (Kavli
IPMU, WPI), University of Tokyo, Chiba 277-8582, Japan}
\altaffiltext{16}{High Energy Accelerator Research Organization and The Graduate University for Advanced Studies, Oho 1-1, Tsukuba, Ibaraki 305-0801, Japan}
\altaffiltext{17}{Academia Sinica Institute of Astronomy and Astrophysics, P.O. Box 23-141, Taipei 10617, Taiwan}

\email{he\_wanqiu@astr.tohoku.ac.jp}
  

\KeyWords{cosmology: large-scale structure of universe --- cosmology: observations --- galaxies: evolution
 --- galaxies: high-redshift --- galaxies: active}

\maketitle

\begin{abstract}

We examine the clustering of quasars over a wide luminosity range, by utilizing 901 quasars at $\overline{z}_{\rm phot}\sim3.8$ with $-24.73<M_{\rm 1450}<-22.23$ photometrically selected from the Hyper Suprime-Cam Subaru Strategic Program (HSC-SSP) S16A Wide2 date release and 342 more luminous quasars at $3.4<z_{\rm spec}<4.6$ having $-28.0<M_{\rm 1450}<-23.95$ from the Sloan Digital Sky Survey (SDSS) that fall in the HSC survey fields. We measure the bias factors of two quasar samples by evaluating the cross-correlation functions (CCFs) between the quasar samples and 25790 bright $z\sim4$ Lyman Break Galaxies (LBGs) in $M_{\rm 1450}<-21.25$ photometrically selected from the HSC dataset. Over an angular scale of \timeform{10.0''} to \timeform{1000.0''}, the bias factors are $5.93^{+1.34}_{-1.43}$ and $2.73^{+2.44}_{-2.55}$ for the low and high luminosity quasars, respectively, indicating no luminosity dependence of quasar clustering at $z\sim4$. It is noted that the bias factor of the luminous quasars estimated by the CCF is smaller than that estimated by the auto-correlation function (ACF) over a similar redshift range, especially on scales below \timeform{40.0''}. Moreover, the bias factor of the less-luminous quasars implies the minimal mass of their host dark matter halos (DMHs) is $0.3$-$2\times10^{12}h^{-1}M_{\odot}$, corresponding to a quasar duty cycle of $0.001$-$0.06$.

\end{abstract}

\section{Introduction}
\label{sec:intro}

It is our current understanding that every massive galaxy is likely to have a super massive black hole (SMBH) at its center \citep{KR1995}. Active Galactic Nuclei (AGNs) are thought to be associated with the growth phase of the BHs through mass accretion. Being the most luminous one of the AGN population, quasars may be the progenitors of the SMBHs in the local universe. Observations over the last decade or so are establishing a series of scaling relations between SMBH mass and properties of their host galaxies (for review see \cite{KHo2013}). A similar scaling relation, involving the mass of the SMBH, is reported even with the host dark matter halo (DMH) mass \citep{Ferrarese2002}. As a result, SMBHs may play an important role in galaxy formation and evolution. However, the physical mechanism behind the scaling relations is still unclear.
 
Clustering analysis of AGNs is commonly used to investigate SMBH growth and galaxy evolution in DMHs. Density peaks in the underlying dark matter distribution are thought to evolve into DMHs (e.g., \cite{PS1974}), in which the entire structure is gravitationally bound with a density 300 times higher than the mean density of the universe. More massive DMHs are formed from rarer density peaks in the early universe, and are more strongly clustered (e.g. \cite{ST1999}; \cite{sheth2001}). If focusing on the large scale clustering, i.e. two-halo term, the mass of quasars host halos can be inferred by estimating the clustering strength of quasars in relative to that of the underlying dark matter, i.e. bias factor. How bias factor of quasars depends on redshift and luminosity provides further information on the relation between SMBHs and galaxies within their shared DMH.

Many studies, based on the two-point correlation function (2PCF) of quasars, have been conducted by utilizing large databases of quasars, such as the 2dF Quasar Redshift Survey (e.g., \cite{croom2005}) and the Sloan Digital Sky Survey (e.g, \cite{Myers2007}; \cite{shen2009}; \cite{white2012}). The redshift evolution of the auto-correlation function (ACF) indicates that quasars are more strongly biased at higher redshifts. For example, luminous SDSS quasars with $-28.2<M_{1450}<-25.8$ at $z\sim4$ show strong clustering with a bias factor of $12.96\pm2.09$, which corresponds to a host DMH mass of $\sim10^{13} h^{-1} M_{\odot}$ \citep{shen2009}. It is suggested that such high luminosity quasar activity needs to be preferentially associated with the most massive DMHs in the early universe \citep{white2008}. If we consider the low number density of such massive DMHs at $z=4$, the fraction of halos with luminous quasar activity is estimated to be 0.03$\sim$0.6 (\cite{shen2007}) or up to 0.1-1 \citep{white2008}. 

The clustering strength of quasars can be also measured from the cross-correlation function (CCF) between quasars and galaxies. When the size of a quasar sample is limited, the clustering strength of the quasars can be constrained with higher accuracy by using the CCF rather than the ACF since galaxies are usually more numerous than quasars. Enhanced clustering and overdensities of galaxies around luminous quasars are expected from the strong auto-correlation of the SDSS quasars at $z\sim4$. However, observational searches for such overdensities around quasars at high redshifts have not been conclusive. While some luminous $z>3$ quasars are found to be in an over-dense region (e.g., \cite{Zheng2006}; \cite{kashikawa2007}; \cite{Utsumi2010}; \cite{Capak2011}; \cite{adams2015}; \cite{GV2017}), a significant fraction of them do not show any surrounding overdensity compared to the field galaxies, and it is suggested that the large scale ($\sim10$ comoving Mpc) environment around the luminous $z>3$ quasars is similar to the Lyman-Break Galaxies (LBGs), i.e. typical star-forming galaxies, in the same redshift range (e.g., \cite{kim2009}; \cite{Hub2013}; \cite{banados2013}; \cite{Uchiyama2017}).

To investigate the quasar environment at $z\sim4$, the clustering of quasars with lower luminosity at $M_{UV}\gtrsim-25$, i.e. typical quasars, which are more abundant than luminous SDSS quasars, is crucial that it can constrain the growth of SMBHs inside galaxies in the early universe \citep{hopkins2007}. At low redshifts ($z\lesssim3$), clustering of quasars is found to have no or weak luminosity dependence (e.g., \cite{Francke2007}; \cite{shen2009}; \cite{krumpe2010}; \cite{Shirasaki2011}). Above $z>3$, \citet{ikeda2015} examined the CCF of 25 less-luminous quasars in the COSMOS field. However, since the sample size is small, the clustering strength of the less-luminous quasars has still not been well constrained, and their correlation with galaxies remains unclear. 

The wide and deep multi-band imaging dataset of the Subaru Hyper Suprime-Cam Strategic Survey Program (HSC-SSP; \cite{aihara2017a}) provides us ac unique opportunity to examine the clustering of galaxies around high-redshift quasars in a wide luminosity range. Based on an early data release of the survey (S16A; \cite{aihara2017b}), a large sample of less-luminous $z\sim4$ quasars ($M_{\rm UV}<-21.5$) is constructed for the first time \citep{akiyama2017}. They cover the luminosity range around the knee of the quasar luminosity function, i.e. they are typical quasars in the redshift range. Additionally, more than 300 SDSS luminous quasars at $z\sim4$ fall within the HSC survey area thanks to a wide filed of 339.8 deg$^2$. Likewise, the five bands of HSC imaging are deep enough to construct a sample of galaxies in the same redshift range through the Lyman-break method \citep{Steidel1996}. 

Here, we examine the clustering of galaxies around $z\sim4$ quasars over a wide luminosity range of $-28.0<M_{\rm 1450}<-22.23$ by utilizing the HSC-SSP dataset. By comparing the clustering of the luminous and less-luminous quasars, we can further evaluate the luminosity dependence of the quasar clustering. The outline of the paper is as follows. Section~\ref{sec:sample} describes the samples of $z\sim4$ quasars and LBGs. Section~\ref{sec:CF} reports the results of the clustering analysis, and we discuss the implication of the observed clustering strength in section~\ref{sec:discussion}. Throughout this paper, we adopt a $\Lambda$CDM model with cosmological parameters of $H_{0}=70$ km s$^{-1}$ Mpc$^{-1}$ ($h=0.7$), $\Omega_{m}=0.3$, $\Omega_{\Lambda}=0.7$ and $\sigma_{8}=0.84$. All magnitudes are described in the AB magnitude system.

\section{Data} 
\label{sec:sample}

\subsection{HSC-SSP Wide-layer dataset}

\label{sec:HSCz4}

We select the candidates of $z\sim4$ quasars and LBGs from the Wide-layer catalog of the HSC-SSP \citep{aihara2017a}. HSC is a wide-field mosaic CCD camera, which is attached to the prime-focus of the Subaru telescope (\cite{miyazaki2012}; \cite{miyazaki2017}). It covers a FoV of 1.5 deg diameter with 116 Full-Depletion CCDs, which have a high sensitivity up to 1$\mu$m. The Wide-layer of the survey is designed to cover 1,400 deg$^{2}$ in the $g$, $r$, $i$, $z$ and $y$ bands with 5 $\sigma$ detection limits of 26.8, 26.4, 26.4, 25.5 and 24.7, respectively, in the 5 year survey \citep{aihara2017a}. In this analysis, we use S16A Wide2 internal data release \citep{aihara2017b}, which covers 339.8 deg$^{2}$ in the 5 bands, including edge regions where the depth is shallower than the final depth. The data are reduced with hscPipe-4.0.2 \citep{Bosch2017}. 

The astrometry of the HSC imaging is calibrated by the Pan-STARRS 1 Processing Version 2 (PS1 PV2) data \citep{Magnier2013}, which covers all HSC survey regions to a reasonable depth with a similar set of bandpasses \citep{aihara2017b}. It is found that the offset RMS of stellar objects between the HSC and PS1 positions is $\sim40$ mas. Extended galaxies have additional offsets with RMS of $\sim30$ mas in relative to the stellar objects \citep{aihara2017b}.

Following the description in sections 2.1 and 2.4 in \citet{akiyama2017}, we construct a sample of objects with reliable photometry (referred as {\it clean} objects hereafter). We apply
\begin{eqnarray}
{\rm flags}\_{\rm pixel}\_{\rm edge} &=& {\rm Not True} \\
{\rm flags}\_{\rm pixel}\_{\rm saturated}\_{\rm center} &=& {\rm Not True} \\
{\rm flags}\_{\rm pixel}\_{\rm cr}\_{\rm center} &=& {\rm Not True} \\
{\rm flags}\_{\rm pixel}\_{\rm bad} &=& {\rm Not True} \\
{\rm detect}\_{\rm is}\_{\rm primary} &=& {\rm True} 
\end{eqnarray}
in all of the 5 bands. These parameters are included as standard output products from the SSP pipeline. 
The (1)-(4) criteria remove objects detected at an edge of the CCDs, affected by saturation within their central 3$\times$3 pixels, affected by cosmic-ray hitting within their central 3$\times$3 pixels and flagged with bad pixels. The final one picks out objects after the deblending process for crowded objects. We apply additional masks (for details see section 2.4 in \cite{akiyama2017}) to remove junk objects. Patches, defined as a minimum unit of a sub-region with an area about \timeform{10.0'} by \timeform{10.0'}, having color offsets in the stellar sequence larger than $0.075$ in either of the $g-r$ vs. $r-i$, $r-i$ vs. $i-z$ and $i-z$ vs. $z-y$ color-color planes are removed (see section 5.8.4 in \cite{aihara2017b}). Tract 8284 is also removed due to unreliable calibration. Moreover, we remove objects close to bright objects by setting the criterion that ${\rm flags}\_{\rm pixel}\_{\rm bright}\_{\rm object}\_{\rm center}$ in all 5 bands are ``Not True''. Regions around objects brighter than 15 in the Guide Star Catalog version 2.3.2 or $i=22$ in the HSC S16A Wide2 database are also removed with masks described in \citet{akiyama2017}. After the masking process, the effective survey area is 172.0 deg$^{2}$.

We use PSF magnitudes for stellar objects and CModel magnitudes for extended objects. PSF magnitudes are determined by fitting a model PSF, while CModel magnitudes are determined by fitting a linear combination of exponential and de Vaucouleurs profiles convolved with the model PSF at the position of each object. We correct for galactic extinction in all 5 bands based on the dust extinction maps by \citet{Schlegel1998}. Only objects that have magnitude errors in the $r$ and $i$ bands smaller than 0.1 are considered.

\subsection{Samples of $z\sim4$ quasars}
\label{sec:HSCz4-qso}

We select candidates of $z\sim4$ quasars from the stellar {\it clean} objects. In order to separate stellar objects from extended objects, we apply the same criteria as described in \citet{akiyama2017},
\begin{eqnarray}
{\rm i}\_{\rm hsm}\_{\rm moments}\_{\rm 11}/{\rm i}\_{\rm hsm}\_{\rm psfmoments}\_{\rm 11} &<& 1.1;  \\
{\rm i}\_{\rm hsm}\_{\rm moments}\_{\rm 22}/{\rm i}\_{\rm hsm}\_{\rm psfmoments}\_{\rm 22} &<& 1.1. 
\end{eqnarray}
${\rm i}\_{\rm hsm}\_{\rm moments}\_{\rm 11} {\rm (22)}$ is the second order adaptive moment of an object in x (y) direction determined with the algorithm described in \citet{HSM2003} and ${\rm i}\_{\rm hsm}\_{\rm psfmoments}\_{\rm 11} {\rm (22)}$ is that of the model PSF at the object position. The $i$-band adaptive moments are adopted since the $i$-band images are selectively taken under good seeing conditions \citep{aihara2017b}. Objects that have the adaptive moment with "nan" are removed. Since stellar objects should have a consistent adaptive moment with that of the model PSF, we set the above stellar/extended clarification criteria. The selection completeness and the contamination are examined by \citet{akiyama2017}. At $i<23.5$, the completeness is above 80\% and the contamination from extended objects is lower than 10\%. At fainter magnitudes ($i>23.5$), the completeness rapidly declines to less than 60\% and the contamination sharply increases to greater than 10\% (see the middle panel of figure 1 in \cite{akiyama2017}). To avoid severe contamination by extended objects, we limit the faint end of the quasar sample to $i=23.5$.

\begin{figure*}
 \begin{center}
   \includegraphics[width=0.45\textwidth]{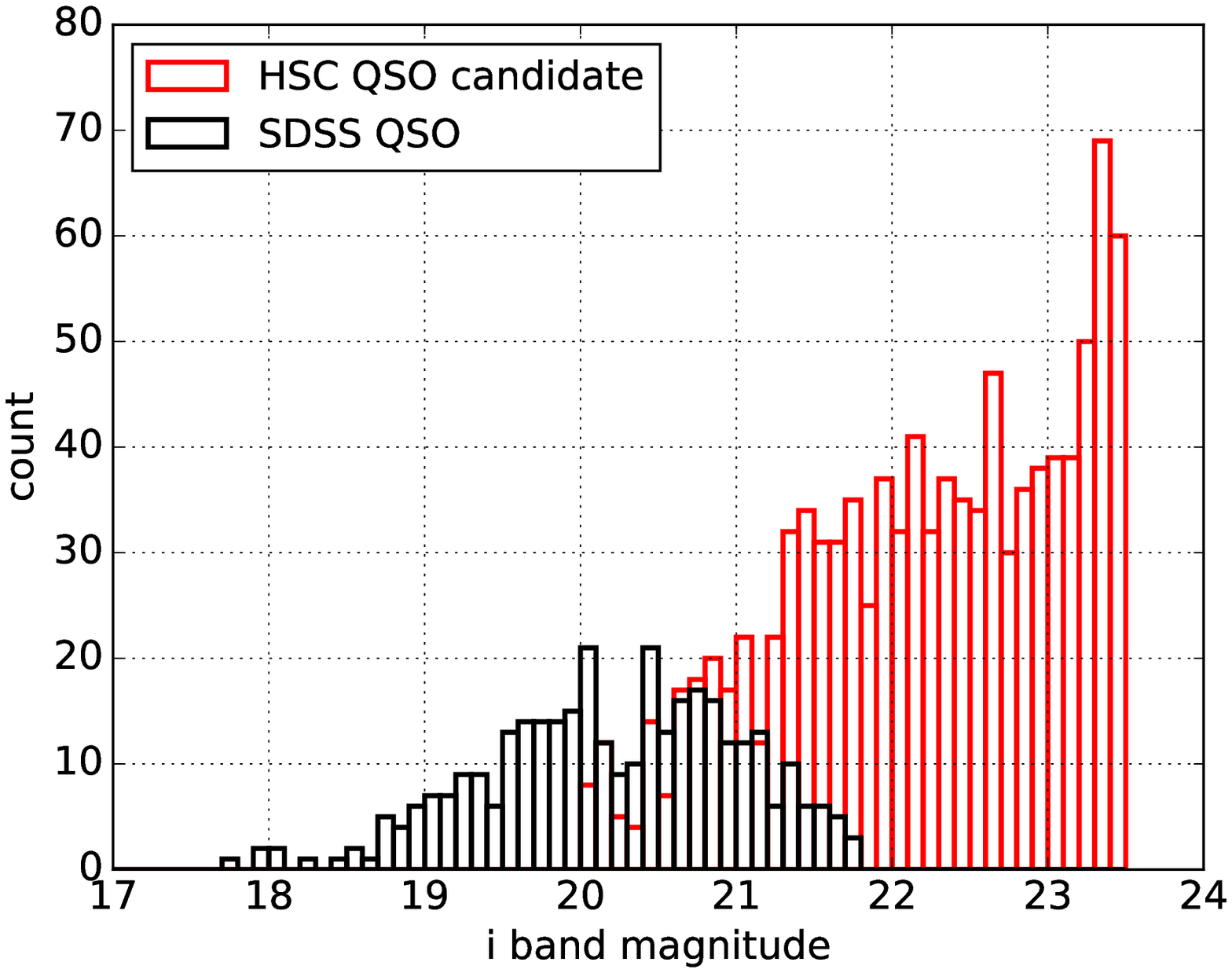}
   \includegraphics[width=0.45\textwidth]{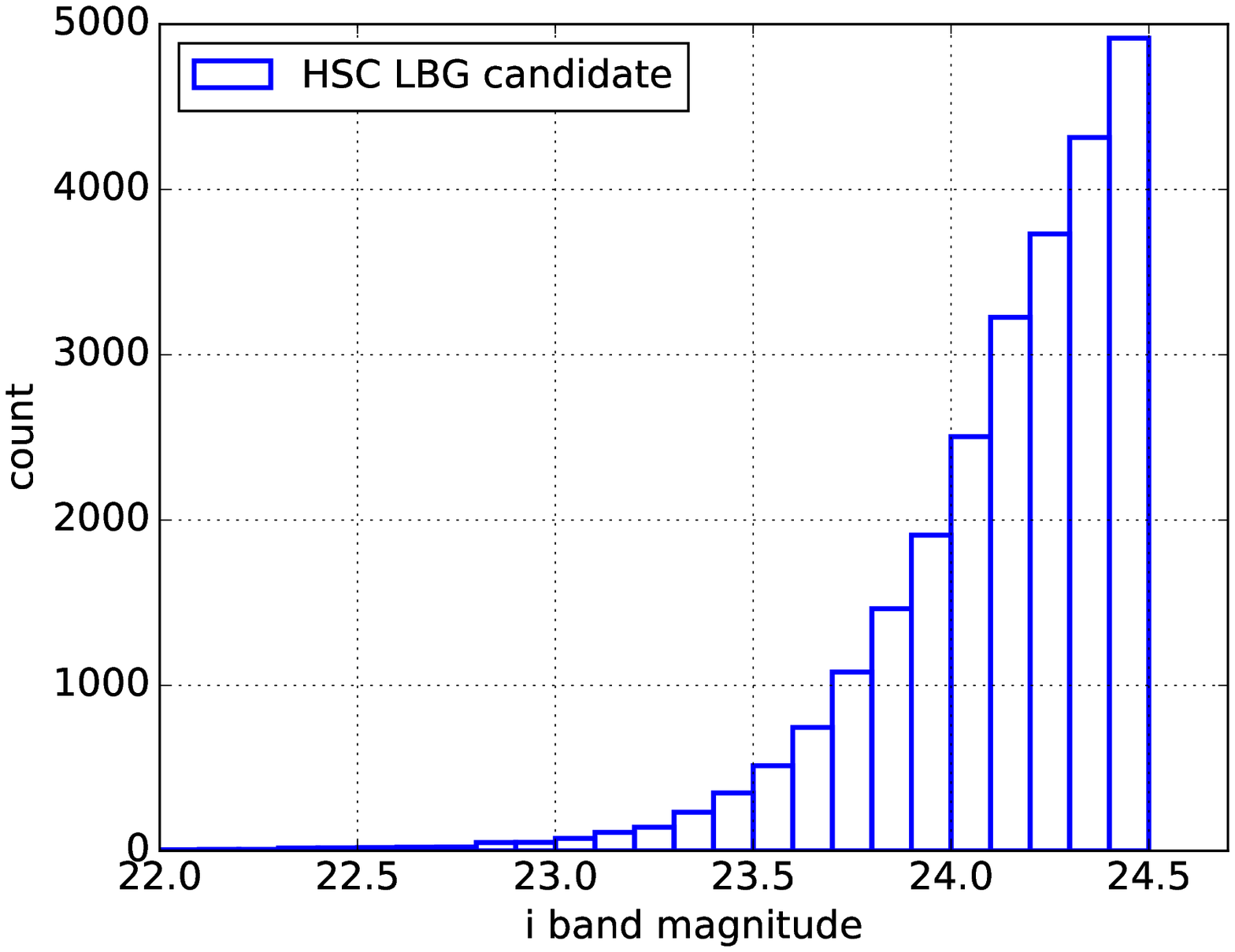}
 \end{center}
 \caption{$i$ band magnitude distributions of the samples. Left: Red and black histograms show the distributions of the $z\sim4$ quasar candidates from the HSC-SSP and SDSS, respectively. Right: Blue histogram represents the distribution of the $z\sim4$ LBGs from the HSC-SSP. A color version of this figure is available in the online journal. }
\label{fig:mag}
\end{figure*}

We apply the Lyman-break selection to identify quasars at $z\sim4$. The selection utilizes the spectral property that the continuum in blue-ward of the Ly$\alpha$ line ($\lambda_{\rm rest}=1216$ {\AA}) is strongly attenuated by absorption due to the intergalactic medium (IGM). The Ly$\alpha$ line of an object at $z=4.0$ is redshifted to 6075 {\AA} in the observed frame, which is in the middle of the $r$-band, as a result the object has a red $g-r$ color. We apply the same color selection criteria as described in \citet{akiyama2017}. In total, 1023 $z\sim4$ quasar candidates in the magnitude range $20.0 <i<23.5$ are selected. We limit the bright end of the sample considering the effects of saturation and non-linearlity. Even though we include edge regions with a shallow depth for the sample selection, we do not find a significant difference of the number densities in the edge and central regions. Therefore, we conclude that larger photometric uncertainties or higher number density of junk objects in the shallower regions do not result in a higher contamination for quasars in the region. The $i$-band magnitude distribution of the sample is shown with the red histogram in the left panel of figure~\ref{fig:mag}.

The completeness of the color selection is examined with the $3.5<z_{spec}<4.5$ SDSS quasars with $i>20.0$ within the HSC coverage \citep{akiyama2017}. Among 92 SDSS quasars with {\it clean} HSC photometry, 61 of them pass the color selection, resulting in the completeness of 66\%. Since the sample is photometrically selected, it can be contaminated by galactic stars and compact galaxies that meet the color selection criteria. The contamination rate is further evaluated by using mock samples of galactic stars and galaxies; the contamination rate is less than 10\% at $i<23.0$, and increases to more than 40\% at $i\sim23.5$. It causes an excess of HSC quasars in faint magnitude bins ($23.2<i<23.5$) as shown in the left panel of figure~\ref{fig:mag}. Since the contamination rate sharply increases at $i>23.5$, we limit the sample at this magnitude. For the bright end, as the luminous SDSS quasar sample primarily includes quasars brighter than $i=21.0$, we consider the HSC quasar sample fainter than $i=21.0$ to constitute the less-luminous quasar sample. Finally, 901 quasars from the HSC are selected in the magnitude range of $21.0<i<23.5$. Here, we convert the $i$-band apparent magnitude to the UV absolute magnitude at 1450 {\AA} using the average quasar SED template provided by \citet{siana2008} at $z\sim4$, which results in a magnitude range of $-24.73<M_{\rm 1450}<-22.23$. In \citet{akiyama2017}, a best fit analytic formula of the contamination rate as a function of the $i$-band magnitude is provided. If we apply it to the less-luminous quasar sample, it is expected that 90 out of 901 candidates are contaminating objects, i.e. contamination rate of the $z\sim4$ less-luminous quasar sample is 10.0\%.

\begin{figure}
 \begin{center}
  \includegraphics[width=0.45\textwidth]{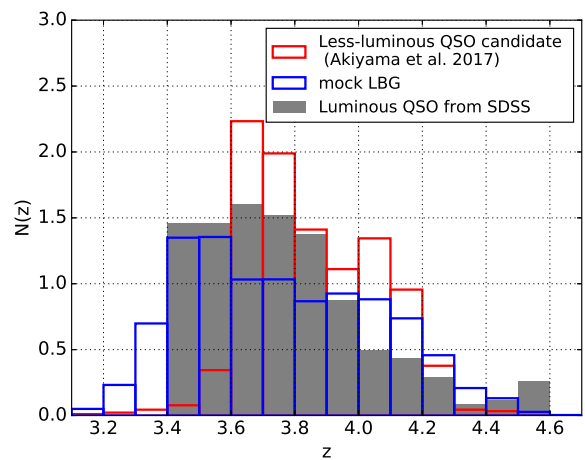}
 \end{center}
\caption{Redshift distributions of the samples. Red histogram indicates the redshift distribution of the less-luminous quasar sample determined either spectroscopically or photometrically \citep{akiyama2017}. Gray filled histogram shows the spectroscopic redshift distribution of the luminous quasar sample. Blue histogram represents the expected redshift distribution of the LBG sample evaluated with the mock LBGs (see text in section~\ref{sec:HSCz4-lbg-z}). All histograms are normalized so that $\int_{0}^{\infty}N(z)dz=1$. A color version of this figure is available in the online journal.}
\label{fig:redshift}
\end{figure}

The redshift distribution of the sample of the $z\sim4$ less-luminous quasar candidates is shown in figure~\ref{fig:redshift} with the red histogram. For 32 candidates with spectroscopic redshift information, we adopt their spectroscopic redshifts, otherwise the redshifts are estimated with a Bayesian photometric redshift estimator using a library of mock quasar templates \citep{akiyama2017}. Most of the quasars are in the redshift range between 3.4 and 4.6. Average and standard deviation of the redshift distribution are $3.8$ and $0.2$, respectively.

In order to examine the luminosity dependence of the quasar clustering, a sample of luminous $z\sim4$ quasars is constructed based on the 12th spectroscopic data release of the Sloan Digital Sky Survey (SDSS) \citep{sdss2015}. We select quasars with criteria on object type (``QSO''), reliability of the spectroscopic redshift (``z\_waring'' flag $=0$), and estimated redshift error (smaller than $0.1$). Only quasars within the coverage of the HSC S16A Wide2 data release are considered. We limit the redshift range between 3.4 and 4.6 following the redshift distribution of the HSC $z\sim4$ LBG sample (which will be discussed in section~\ref{sec:HSCz4-lbg-z}). In the coverage of the HSC S16A Wide2 data release, there are 342 quasars that meet the selection criteria. Their redshift distribution is shown by gray filled histogram in figure~\ref{fig:redshift}. Average and standard deviation of the redshift distribution are $3.77$ and $0.26$, respectively. Although the redshift distribution of the SDSS sample shows excess around $z\sim3.5$ compared to the HSC sample, the average and standard deviations are close to each other.

The $i$-band magnitude distribution of the SDSS quasars is plotted by the black histogram in the left panel of figure~\ref{fig:mag}. To determine their $i$-band magnitude in the HSC photometric system, we match the sample to HSC {\it clean} objects using a search radius of \timeform{1.0''}. Out of the 342 SDSS quasars, 296 have a corresponding object among the {\it clean} objects, while the others are saturated in the HSC imaging data. For the remaining 46 quasars, we convert their $r$- and $i$-band magnitudes in the SDSS system to the $i$-band magnitude in the HSC system following the equations in section 3.3 in \citet{akiyama2017}. As can be seen from the distributions, the SDSS quasar sample covers a magnitude range about 2 magnitude brighter than the HSC quasar sample. Their corresponding UV absolute magnitudes at 1450 {\AA} are in the range of $-28.0$ to $-23.95$ evaluated by the same method with the less-luminous quasar sample.

\subsection{Sample of $z\sim4$ LBGs from the HSC dataset}
\label{sec:HSCz4-lbg}

We select candidates of $z\sim4$ LBGs from the S16A Wide2 dataset in the similar way as we select the $z\sim4$ quasar candidates. Different from the quasars, we select candidates from the extended {\it clean} objects instead of the stellar objects, i.e. we pick out the {\it clean} objects that do not meet either of the equations (6)(7) as extended objects. As shown in figure 9 of \citet{akiyama2017}, extended galaxies at $z>3$ are distinguishable from stellar quasars with these criteria, as a result of the good image quality of the $i$-band HSC Wide-layer images, which has a median seeing size of \timeform{0.61''} \citep{aihara2017b}. While the stellar/extended classification is ineffective at $i>23.5$, the contamination of stellar objects to the LBG sample is negligible, because the extended objects outnumber the stellar objects by $\sim30$ times at $23.5<i<25.0$.

\begin{figure}
 \begin{center}
  \includegraphics[width=0.45\textwidth]{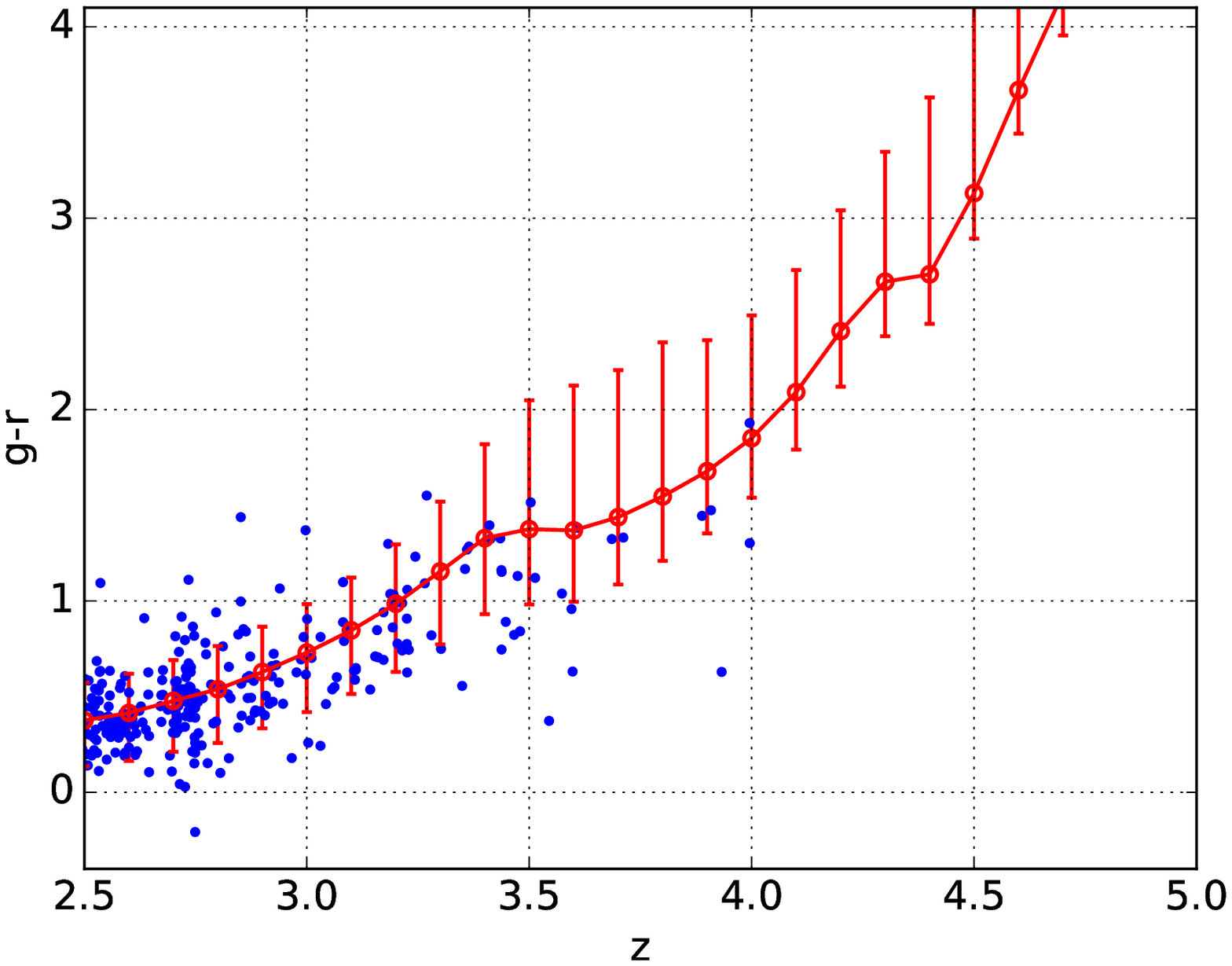}
  \includegraphics[width=0.45\textwidth]{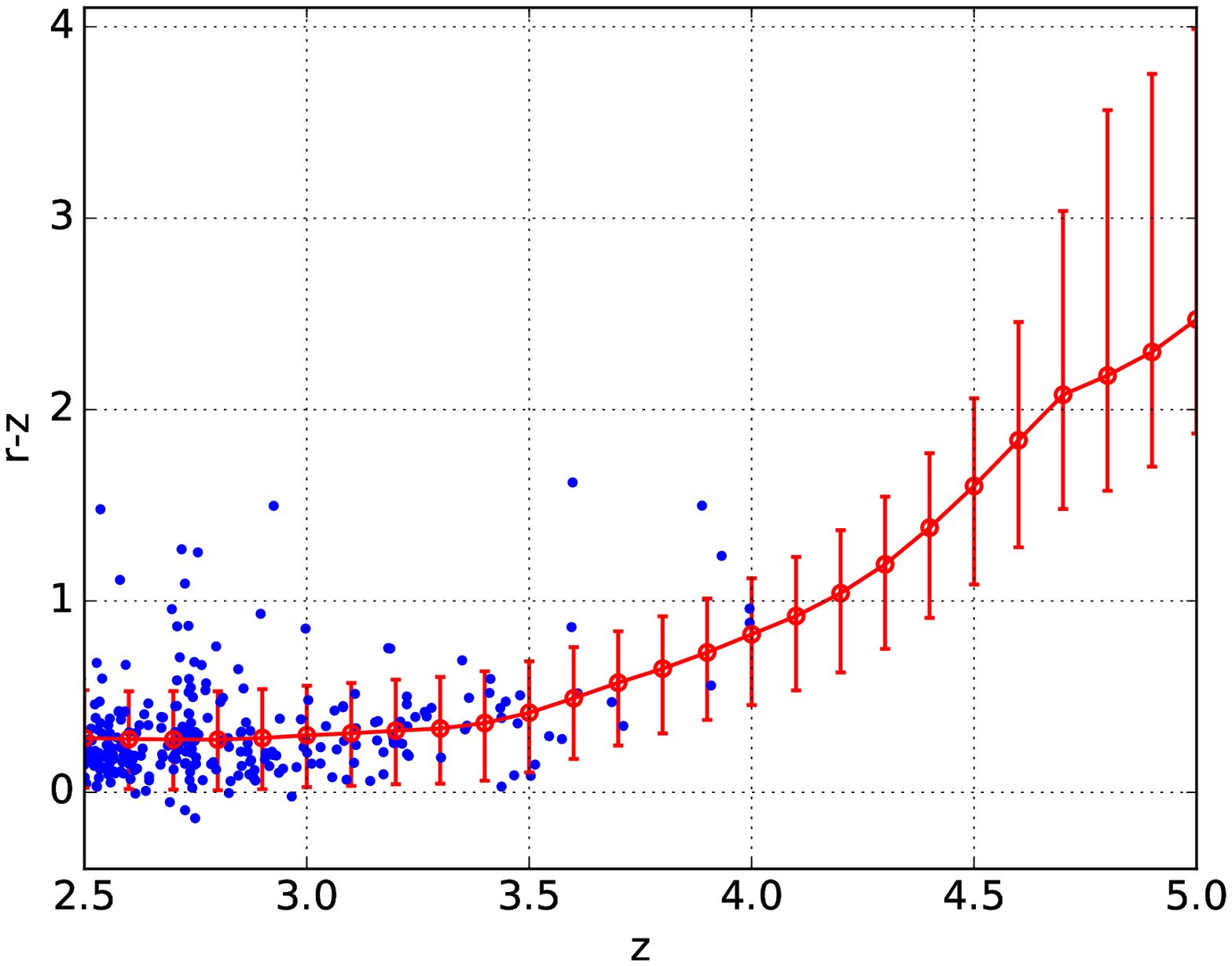}
 \end{center}
 \caption{$g-r$ (top) and $r-z$ (bottom) colors versus redshift of the mock LBGs. The red line and the error bars are the average and 1$\sigma$ scatter of the colors of the mock LBGs. Blue points represent spectroscopically confirmed galaxies within the HSC S16A Ultra-Deep layer.
}\label{fig:LBG-ccd}
\end{figure}

We determine the color selection criteria of $z\sim4$ LBGs based on color distributions of a library of model LBG spectral energy distributions (SEDs), because the sample of $z\sim4$ LBGs with a spectroscopic redshift at the depth of the HSC Wide-layer is limited. The model SEDs are constructed with the stellar population synthesis model by \citet{BC2003}. We assume a Salpeter initial mass function \citep{salp1955} and the Padova evolutionary track for stars (\cite{Fagotto1994a}; \cite{Fagotto1994b}) of solar metallicity. Following a typical star-formation history of $z\sim4$ LBGs derived based on an optical-NIR SED analysis (e.g. \cite{Shapley2001}; \cite{nonino2009}; \cite{Yabe2009}), we adopt an exponentially declining star-formation history with $\psi(t)=\tau^{-1}$exp$(-t/\tau)$, where $\tau=$50 Myr and $t=300$ Myr. In addition to the stellar continuum component, we also consider the Ly$\alpha$ emission line at 1216 {\AA} with a EW$_{Ly\alpha}$ randomly distributed within the range between 0 and 30 {\AA}, which is determined to follow the Ly$\alpha$ EW distribution of luminous LBGs in the UV absolute magnitude range of $-23.0$ $\sim$ $-21.5$ \citep{ando2006}. We apply extinction as a screen dust with the dust extinction curve of \citet{calzetti2000}. We assume that E(B$-$V) has a Gaussian distribution with a mean of 0.14 and 1$\sigma$ of 0.07 following that observed for $z\sim3$ UV-selected galaxies \citep{reddy2008}. In order to reproduce the observed scatter of the $g-r$ color of galaxies at $z\sim3$ (see figure~\ref{fig:LBG-ccd}), the scatter of the color excess is doubled to $\sigma=0.14$. In total, 3,000 SED templates are constructed. Each template is redshifted to $z=2.5$-$5.0$ with an interval of 0.1. Attenuation by the intergalactic medium is applied to the redshifted templates. We follow the updated number density of the Ly$\alpha$ absorption systems in \citet{inoue2014}, and consider scatter in the number density of the systems along different line of sights with the Monte Carlo method used in \citet{inoue2008} (Inoue, private communication). In Figure~\ref{fig:LBG-ccd}, we compare the distributions of the $g-r$ and $r-z$ colors of the templates with those of the galaxies with spectroscopic redshift in the HSC-SSP catalogs of the Ultra-Deep layer. The color distribution of the mock LBGs as a function of redshift reproduces that of the galaxies with spectroscopic redshifts around 3. At $z>3.5$, it is hard to judge the consistency due to the limited size of galaxies with available spectroscopic redshifts.

\begin{figure}
 \begin{center}
  \includegraphics[width=0.5\textwidth]{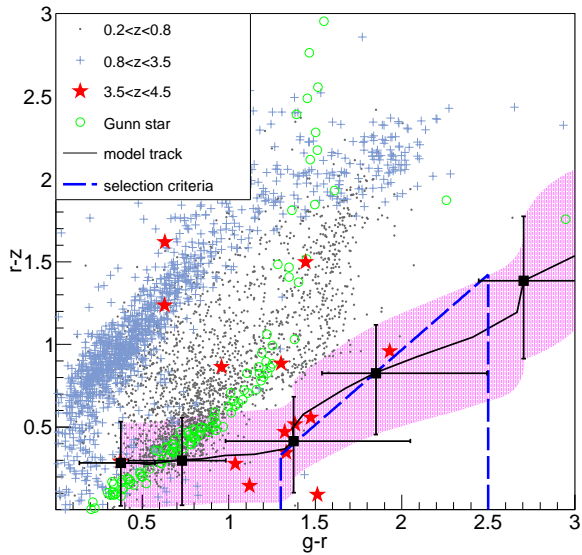}
 \end{center}
 \caption{Color selection of z$\sim$4 LBGs. Gray points and blue crosses are galaxies at $0.2<z<0.8$ and $0.8<z<3.5$, respectively. Only 5.0\% of them are plotted for the clarification. Red stars are galaxies at $3.5<z<4.5$. Green open circles are colors of stars derived in the spectro-photometric catalog by \citet{Gunn1983}. Solid black line is the track of the model LBG. Black squares and error bars denote the average and 1$\sigma$ color scatter of the mock LBGs along the track at $z=$2.5, 3.0, 3.5, 4.0 and 4.4. Pink shaded area implies the 1$\sigma$ $r-z$ scatter of the mock LBGs. Blue dashed lines represent our selection criteria. A color version of this figure is available in the online journal.}
\label{fig:LBG-CCD}
\end{figure}

Considering the color distributions of the mock LBGs and the LBGs with a spectroscopic redshift, we determine the color selection criteria on the $g-r$ vs. $r-z$ color-color diagram as shown in figure~\ref{fig:LBG-CCD} with the blue dashed lines. Gray dots and blue crosses represent colors of galaxies with a spectroscopic redshift at $0.2<z<0.8$ and $0.8<z<3.5$, respectively, in the HSC Wide-layer photometry. Red stars are galaxies at $3.5<z<4.5$. We plot the color track of the model LBG with the black solid line, and mark the colors at $z=2.5, 3.0, 3.5, 4.0$ and $4.4$ with the $1\sigma$ scatter. The pink shaded region represents 1$\sigma$ scatter of the $r-z$ color along the model track. The selection criteria are
\begin{eqnarray}
  0.909(g-r)-0.85&>&(r-z);\\
  (g-r)&>&1.3;\\
  (g-r)&<&2.5.
\end{eqnarray}
We determine the selection criteria to enclose the large part of the color distribution of the models while preventing severe contamination from low-redshift galaxies. The third criterion limits the upper redshift range of the sample, and is adjusted to match the expected redshift distribution of the less-luminous $z\sim4$ quasars. In order to reduce contaminations by low-redshift red galaxies and objects with unreliable photometry, we consider two additional criteria 
\begin{eqnarray}
  (i-z)&<&0.2;\\
  (z-y)&<&0.2
\end{eqnarray}
following figure 3 in \citet{akiyama2017}. Because the contamination by low-redshift galaxies is severe at magnitude fainter than $i=24.5$, we limit the sample at this magnitude. Finally, we select 25790 $z\sim4$ LBG candidates at $i<24.5$. The $i$-band magnitude distribution of the candidates is shown in the right panel of figure~\ref{fig:mag}. The brightest candidate is at $i=21.87$, but there are only 4 candidates at $i<22$. Thus we plot the distribution from $i=22$. The corresponding UV absolute magnitudes of the candidates at 1450 {\AA} are evaluated to be in the range of $-23.88<M_{\rm 1450}<-21.25$ by the model LBG at $z\sim4$. It should be noted that there is a difference in the sky coverage between both of the quasar samples and $i<24.5$ LBGs, because of the edge regions with shallow depth where only the quasars are selected reliably. Such selection effects are taken into consideration when constructing the random sample (section~\ref{sec:HSCz4-rand}).

\subsection{Redshift distribution and contamination rate of the $z\sim4$ LBG sample}
\label{sec:HSCz4-lbg-z}

The redshift distribution of the LBG sample is evaluated by applying the same selection criteria to a sample of mock LBGs, which are constructed in the redshift range between 3.0 and 5.0 with a 0.1 redshift bin. At each redshift bin, we randomly select LBG templates from our library of SEDs and normalize them to have $22.0<i<24.5$ following the LBG UV luminosity function at $z\sim3.8$ \citep{Burg2010}. We convert the apparent $i$-band magnitude to the absolute UV magnitude based on the selected templates. It should be noted that an object with a fixed apparent magnitude has a higher luminosity and a smaller number density in the luminosity function at higher redshifts. We also consider the difference in comoving volume at each redshift bin. 

For each redshift bin, we then place the mock LBGs at random positions in the HSC Wide-layer images with a density of 2,000 galaxies per deg$^{2}$, and apply the same masking process as for the real objects. We calculate the expected photometric error at each position using the relation between the flux uncertainty and the value of image variance. This relation is determined empirically with the flux uncertainty of real objects as a function of the PSF and object size. The variance is measured within \timeform{1''}$\times$\timeform{1''} at each point. The size of the model PSF at the position is evaluated with the model PSF of the nearest real object in the database. In order to reproduce the photometric error associated with the real LBGs, we use the relation for a size of \timeform{1.5''}. After calculating the photometric error with this method, we add a random photometric error assuming the Gaussian distribution. Finally, we apply the color selection criteria and remove mock LBGs with the magnitude error in either of $i$- or $r$- band larger than 0.1. The ratio of the recovered mock LBGs to the full random mock LBGs is evaluated as the selection completeness at each redshift bin. We find that the selection completeness is $\sim10.0-30.0$\% in the redshift range between 3.5 and 4.2, but smaller than 5\% at other redshifts. These low rates are due to the fact that we set stringent constraints so that we can prevent the severe contamination from low-redshift galaxies. Based on a selection completeness of 20.0\% at $3.5<z<4.2$, we calculate an expected number of 35988 LBGs with $22<i<24.5$ in the HSC-SSP S16A Wide-layer from the LBG UV luminosity function at $z\sim3.8$ \citep{Burg2010}, which is larger than the actual LBG sample size (25790) in this work since we consider the edge regions that have a shallow depth. The effect of the shallow depth is considered in the construction of the random objects (section~\ref{sec:HSCz4-rand}).

The redshift distribution is measured by multiplying the completeness ratio with the number of mock LBGs at each redshift, which is shown in figure~\ref{fig:redshift} with the blue histogram. The average and 1$\sigma$ of the distribution is 3.71 and 0.30, respectively. The redshift distribution of the LBGs is similar to that of the luminous quasar sample, but slightly extended toward lower redshifts than the less-luminous quasar sample. It is likely that the extension is due to the higher number density of LBGs in $22.0<i<24.5$ at $3.3<z<3.5$. 

The LBG sample can be contaminated by low-redshift red galaxies which have similar photometric properties to the $z\sim4$ LBGs. We evaluate the contamination rate of the LBG selection using the HSC photometry in the COSMOS region and the COSMOS $i$-band selected photometric redshift catalogue, which is constructed by a $\chi^{2}$ template-fitting method with 30 broad, intermediate, and narrow bands from UV to mid-IR in the 2-deg$^{2}$ COSMOS field \citep{Ilbert2008}. In the HSC-SSP S15B internal database, three stacked images in the COSMOS region, simulating good, median, and bad seeing conditions, are provided. Since the $i$-band images of the Wide-layer are selectively taken under good or median seeing conditions \citep{aihara2017b}, we match the catalogs from the median stacked image, which has a FWHM of \timeform{0.70''}, with galaxies in the photometric redshift catalog within an angular separation of \timeform{1.0''}. As examined by \citet{Ilbert2008}, the photometric redshift uncertainty of galaxies with the COSMOS $i'-$band magnitude brighter than $24.0$ is estimated to be smaller than $0.02$ at $z<1.25$. For galaxies within the same luminosity range at higher redshifts $1.25<z<3$, the uncertainty is significantly higher but roughly below $0.1$. Thus we only include objects with photometric redshift uncertainty less than 0.02 and 0.1 at $z<1.25$ and at $z>1.25$, respectively, in the matched catalog. We apply the color selection criteria (8)-(12) to the matched catalog. Among 700 matched galaxies with $3.5<z_{phot}<4.5$, 117 galaxies pass the selection criteria, resulting in the completeness of 17\%, which is consistent with that examined by the mock LBGs. Meanwhile, we investigate the contamination by the ratio of galaxies at $z<3$ or $z>5$ among those passing the selection criteria at each magnitude bin of 0.1. It is found that the contamination rate is 10\% to 30\% in the magnitude range of $i=23.5-24.5$, and sharply increases to $>$50\% at $i=25.0$. In total, all contaminating sources are clarified to be at $z<3$, while 95\% of them are at $z<1$. We multiply the contamination rate as a function of the $i$-band magnitude with the number counts of the LBG candidates at each $0.1$ bin to estimate the total number of contaminating sources in the sample. Among 25790 LBG candidates, 5886 are expected to be contaminating objects at $z<3$, i.e. the contamination rate is 22.8\%.

Furthermore, we also check the photometric redshift of the LBG candidates determined with the 5 bands HSC Wide-layer photometry by the MIZUKI photometric redshift code, which uses the Bayesian photometric redshift estimation \citep{MIZUKI}. Among the 25790 $z\sim4$ LBG candidates, 25749 of them have photometric redshift with the MIZUKI code, and 4091 of them have photometric redshift lower than $z=3.0$. The contamination rate is evaluated to be 15.9\%, which is similar to the one evaluated in the COSMOS region. Since the COSMOS photometric redshift catalog is based on the 30 bands photometry covering wider wavelength coverage, we consider the contamination rate evaluated in the COSMOS region in the later clustering analysis. 

\subsection{Constructing random objects for the clustering analysis}
\label{sec:HSCz4-rand}

The clustering strength is evaluated by comparing the number of pairs of real objects and that of mock objects distributed randomly in the survey area. Therefore it is necessary to construct a sample of mock objects that are distributed randomly within the survey area and are selected with the same selection function as the real sample. From $z=3$ to 5, we construct 3000 mock LBG SEDs, which are normalized to have $i=24.5$, at each 0.1 redshift bin. Then we place the mock LBGs randomly over the survey region with the surface number density of 2,000 LBGs per deg$^{2}$ with errors as described in section~\ref{sec:HSCz4-lbg-z}. After applying the same color selection and magnitude error criteria as for the real objects, we create a sample of 150,756 random LBGs, which reproduce the global distribution of the real LBGs including the edge of the survey region where the depth is shallower. Therefore, the clustering analysis is not affected by the discrepancy of the sky coverage between the quasars and LBGs.

\section{Clustering analysis}
\label{sec:CF}

\subsection{Cross-correlation functions of the less-luminous and luminous quasars at $z\sim4$}
\label{sec:CCF}

\begin{figure*}[!h]
 \begin{center}
   \includegraphics[width=0.45\textwidth]{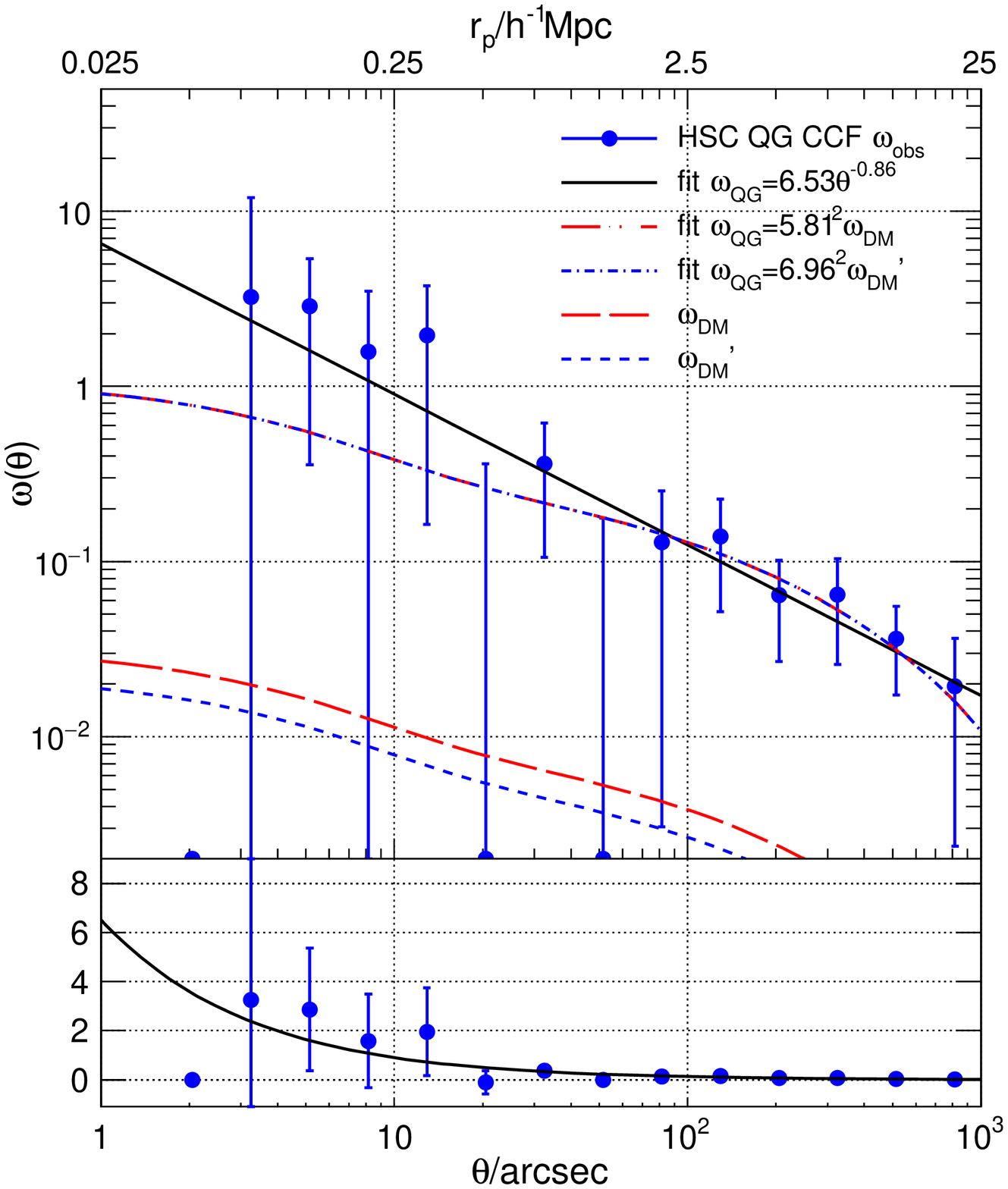}
   \includegraphics[width=0.45\textwidth]{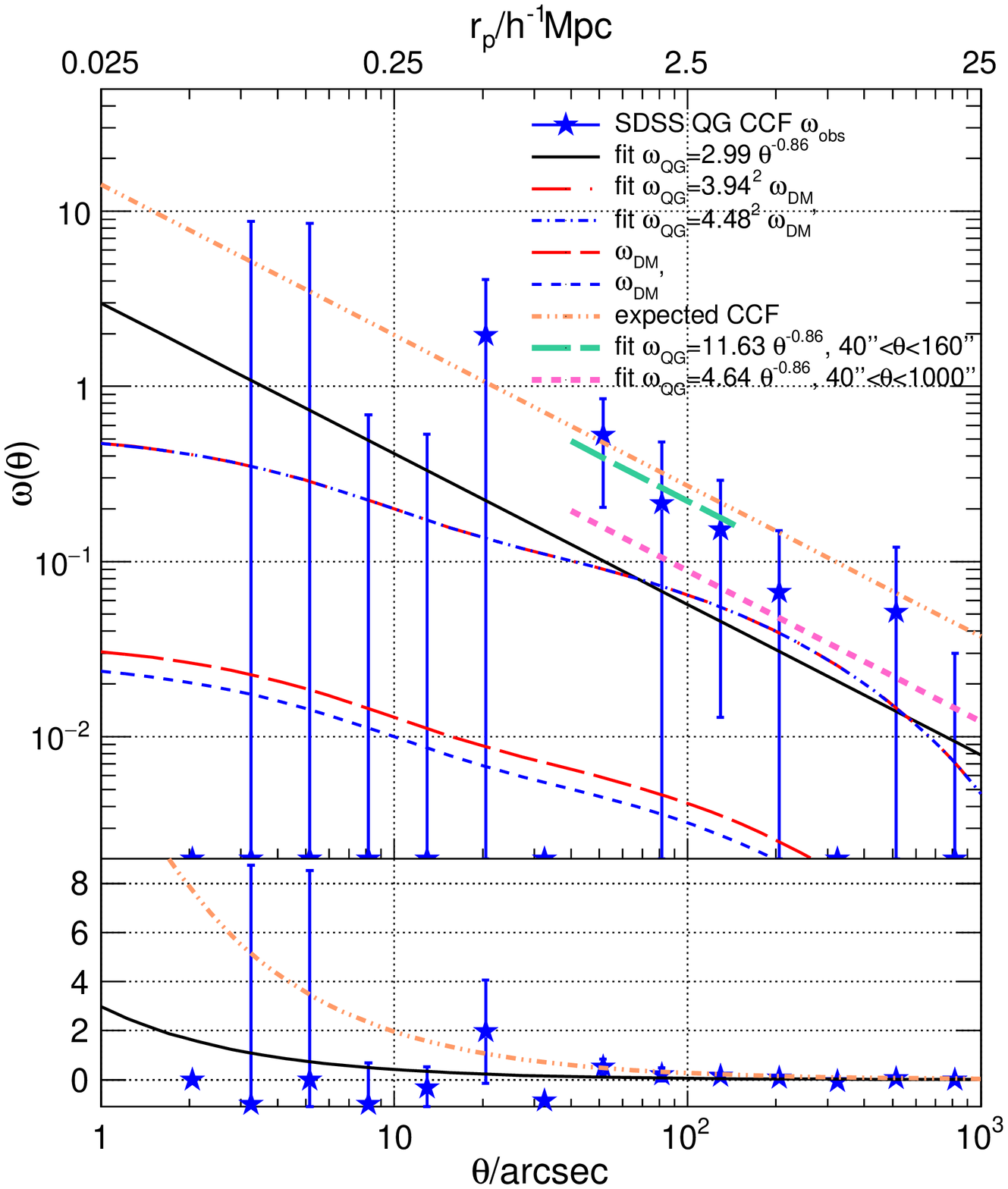}
 \end{center}
 \caption{Left panel: blue dots are the observed CCF $\omega_{\rm obs}$ of the less-luminous quasars and the LBGs at $z\sim4$. Black solid line is the best fit power-law model using ML fitting in the scale of \timeform{10.0''} to \timeform{1000.0''}. Red dash-dotted line is the best fit dark matter model $\omega_{\rm DM}$ (red long-dashed line) adopting ML fitting in the same scale based on the HALOFIT power spectrum \citep{smith2003}, while the blue dash-dotted line is the best fit dark matter model $\omega_{\rm DM}'$ (blue short-dashed line) after considering the contaminations of the less-luminous quasars and the LBGs. Right panel: blue stars are the observed CCF $\omega_{\rm obs}$ of the luminous quasars and the LBGs at $z\sim4$. Red and blue lines have the same meaning with the the left panel but blue line only considers the contamination of the LBGs. Orange dash-dot-dotted line is the expected CCF of the luminous quasars estimated by the luminous quasars ACF in \citet{shen2009}. Green thick long-dashed and pink thick dashed line are the best fit power-law models in the scale of \timeform{40.0''} to \timeform{160.0''} and of \timeform{40.0''} to \timeform{1000.0''}, receptively. In both of the panels, symbols just on the horizontal axis with no error bar beyond $\timeform{10.0''}$, with no error bar within \timeform{10.0''} and with error bars in the top pad mean negative bins with a small error bar, zero bins without pair count and negative or zero bins with a large error bar. Top and bottom panels show the logarithmic and the linear scale of the vertical axis respectively. Top horizontal axis of the top panel implies the comoving distance at redshift 4. 
}\label{fig:CCF}
\end{figure*}

We evaluate the CCFs of the $z\sim4$ quasars and LBGs with the projected two point angular correlation function, $\omega(\theta)$, since most of the quasar and LBG candidates do not have spectroscopic redshifts. We use the estimator from \citet{DP1983},
\begin{equation}
\omega(\theta)=\frac{DD(\theta)}{DR(\theta)}-1,
\end{equation}
where $DD(\theta)=\langle DD \rangle/N_{\rm QSO}N_{\rm LBG}$ and $DR(\theta)=\langle DR \rangle/N_{\rm QSO}N_{\rm R}$ are the normalized quasar - LBG pair counts and quasar - random LBG pair counts in an annulus between $\theta-\Delta \theta$ and $\theta+\Delta \theta$, respectively. Here, $\langle DD \rangle$ and $\langle DR \rangle$ are the numbers of quasar - LBG and quasar - random LBG pairs in the annulus, and $N_{\rm QSO}$,  $N_{\rm LBG}$ and $N_{\rm R}$ are the total numbers of quasars, LBGs and random LBGs, respectively. We set 14 bins from \timeform{1.0"} to \timeform{1000.0"} in the logarithmic scale. The CCFs of the quasars and LBGs for the less-luminous and luminous quasars are plotted in the left and right panels of figure~\ref{fig:CCF}, respectively, and summarized in table~\ref{tab:CCF} along with the pair count in each bin. 
\begin{table*}
 \tbl{The less-luminous and luminous quasar - LBG CCFs at z$\sim$4}{
  \begin{tabular}{ccccccccccc}
     \hline
     & & &  less-luminous & & & &luminous& &\\
     $\theta$(arcsec) &($\theta_{\rm min}$, $\theta_{\rm max}$) &$\langle D_{Q}D_{G} \rangle$& $\langle D_{Q}R_{G} \rangle$ & $\omega(\theta)$ & $\sigma(\theta)$&$\langle D_{Q}D_{G} \rangle$& $\langle D_{Q}R_{G} \rangle$ & $\omega(\theta)$ & $\sigma(\theta)$  \\
\hline
2.05 &(1.58, 2.51) & 0 & 0 & 0 & 0  & 0 & 0 & 0 & 0\\
3.25 &  (2.51, 3.98) &2 & 3 & 3.25 & 8.74& 0 & 1 & -1 & 9.76\\
5.15 &(3.98, 6.31) & 4 & 6 & 2.86& 2.51&3 & 0 & 0 & 8.53\\
8.15 & (6.31, 10.00) &3 & 7 & 1.58 & 1.91&0 & 4 & -1 & 1.69\\
12.92 &(10.00, 15.85) &5 & 10 & 1.96 & 1.79& 1 & 9 & -0.33 & 0.86\\
20.48 & (15.85, 25.12) &7 & 47 & -0.11 & 0.47& 3 & 6 & 1.96 & 2.11\\
32.46 & (25.12, 39.81) &28& 120 & 0.36 & 0.26& 1 & 41 & -0.85 & 0.15\\
51.45 & (39.81, 63.10) &52 & 303 & -0.002 & 0.18& 25 & 96 & 0.53 & 0.32\\
81.55 &  (63.10, 100.00) &143 & 739 & 0.13 & 0.13& 47 & 226 & 0.21 & 0.27\\
129.24 &(100.00, 158.49) & 334 & 1710 & 0.14 & 0.09& 116 & 589 & 0.15 & 0.14\\
204.84 & (158.49, 251.19) & 754 & 4144 & 0.06 & 0.04& 257 & 1407 & 0.07 & 0.08\\
324.65 & (251.19, 398.11) & 1887 & 10375 & 0.06 & 0.04& 585 & 3677 & -0.07 & 0.04\\
514.53 & (398.11, 630.96) & 4564 & 25764 & 0.04 & 0.02& 1669 & 9272 & 0.05 & 0.07\\
815.48 & (630.96, 1000.00) &11065 & 63358 & 0.02 & 0.02& 3967 & 23241 & -0.005 & 0.03\\
\hline
  \end{tabular}}\label{tab:CCF}
\end{table*}

The uncertainty of the CCFs is evaluated through the Jackknife resampling \citep{Zehavi2005}. We separate the survey area into $N=22$ subregions with a similar size. In $i$-th resampling, we ignore one of the subregions to construct a new set of samples of quasars, LBGs, and random LBGs and estimate their correlation function, $\omega_{i}$. We evaluate the uncertainty only by the diagonal elements of the covariance matrix
\begin{equation}
{\rm Cov}(\omega_{i},\omega_{j})=\frac{N-1}{N}{\sum_{k=1}^{N}(\omega^{k}_{i}-\overline{\omega_{i}})(\omega^{k}_{j}-\overline{\omega_{j}})},
\end{equation}
where $\overline{\omega_{i}}$ is the mean of $\omega_{i}$ over the $N$ Jackknife samples, because the diagonal elements are sufficienct to recover the true uncertainty \citep{Zehavi2005}. The $\overline{\omega_{i}}$ at each radius bin is consistent with the CCFs of the whole samples of the less-luminous and luminous quasars. The resulting uncertainty with the Jackknife resampling is about 1.5 - 2 times larger than the Poisson error ($\sigma(\theta)=(1+\omega(\theta))/\sqrt{N_{pair}}$) in the scale beyond \timeform{500.0''}. But these two error estimators are consistent with each other in the scale within \timeform{300.0''}. In the scale smaller than \timeform{20.0''}, due to the limited quasar-LBG pair count, the Poisson error can be even larger than the Jackknife one if we evaluate the Poisson uncertainty with the Poisson statistics for a small sample \citep{neil1986}. Here, since we do not consider the small scale within \timeform{10.0''} in the fitting process, we adopt the Jackknife error for the CCF beyond \timeform{10.0''}. For the scale within \timeform{10.0''}, if the Jackknife estimator fails to give a value due to either of no $\langle DD \rangle$ or $\langle DR \rangle$ pair count in any subsamples, we show the Poisson error following the Poisson statistics for a small sample \citep{neil1986} in table~\ref{tab:CCF} and figure~\ref{fig:CCF}. 

\begin{table*}[h!]
 \tbl{Summary of clustering analysis for the CCFs}{%
 \begin{tabular}{cccccccccc}
 \hline
     CF &model&fitting& $\bar{z}$ & [$\theta_{min},\theta_{max}$] & $A_{\omega}$ & $r_{0}$    &$b_{QG}$   &  $b_{QSO}$ & ${\rm log} M_{\rm DMH}$ \\
        &     &       &           & (arcsec)                     &             &($h^{-1}$ Mpc) &      & &($h^{-1}M_{\odot}$)\\
 \hline
       & power-law & $\chi^{2}$ & 3.80 & [10 , 1000] & $6.03^{+1.65}_{-1.65}$ & $7.13^{+0.99}_{-1.13}$ & $5.62^{+0.72}_{-0.82}$& $5.48^{+1.25}_{-1.32}$&      $12.07^{+0.33}_{-0.49}$         \\
       & power-law$'$ $*$  & $\chi^{2}$ & 3.80 & [10 , 1000] & $8.67^{+2.37}_{-2.37}$ & $8.66^{+1.20}_{-1.37}$ & $6.74^{+0.87}_{-0.98}$& $6.10^{+1.40}_{-1.47}$&      $12.25^{+0.32}_{-0.47}$         \\
  Less-   & power-law & ML & 3.80 & [10 , 1000] & $6.53^{+1.85}_{-1.81}$ & $7.44^{+1.07}_{-1.19}$ & $5.85^{+0.78}_{-0.87}$&$5.94^{+1.42}_{-1.46}$& $12.20^{+0.33}_{-0.49}$     \\
luminous     & power-law$'$  & ML & 3.80 & [10 , 1000] & $9.39^{+2.66}_{-2.60}$ & $9.04^{+1.30}_{-1.45}$ & $7.01^{+0.93}_{-1.04}$&$6.60^{+1.57}_{-1.63}$& $12.37^{+0.32}_{-0.47}$     \\
QG    & DM & $\chi^{2}$  & 3.80 & [10 , 1000] & -& - & $5.68^{+0.70}_{-0.80}$&$5.67^{+1.23}_{-1.32}$& $12.13^{+0.31}_{-0.46}$          \\
 CCF   & DM$'$  & $\chi^{2}$  & 3.80 & [10 , 1000] & -& - & $6.76^{+0.83}_{-0.94}$&$6.21^{+1.34}_{-1.42}$& $12.28^{+0.30}_{-0.44}$          \\
       & DM & ML  & 3.80 & [10 , 1000] &- & - & $5.81^{+0.74}_{-0.85}$&$5.93^{+1.34}_{-1.43}$& $12.20^{+0.32}_{-0.48}$       \\
       & DM$'$ & ML  & 3.80 & [10 , 1000] &- & - & $6.96^{+0.89}_{-1.01}$&$6.58^{+1.49}_{-1.58}$& $12.37^{+0.31}_{-0.45}$       \\
    \hline
     & power-law & ML & 3.77 & [10 , 1000] & $2.99^{+3.08}_{-2.97}$ & $4.73^{+2.19}_{-4.41}$ & $3.77^{+1.60}_{-3.19}$&$2.47^{+2.36}_{-2.41}$&   $10.45^{+1.40}_{-10.45}$        \\
     & power-law$'$ & ML & 3.77 & [10 , 1000] & $3.87^{+3.98}_{-3.84}$ & $5.43^{+2.52}_{-5.06}$ & $4.29^{+1.82}_{-3.63}$&$2.47^{+2.37}_{-2.41}$& -       \\
     & power-law & ML & 3.77 & [40 , 160] & $11.63^{+6.55}_{-6.07}$ & $9.81^{+2.66}_{-3.32}$ & $7.44^{+1.86}_{-2.24}$&$9.61^{+4.88}_{-4.73}$& $12.92^{+0.53}_{-1.05}$       \\
Luminous     & power-law & ML & 3.77 & [40 , 1000] & $4.64^{+3.27}_{-3.20}$ & $5.99^{+1.99}_{-2.80}$ & $4.70^{+1.44}_{-2.01}$&$3.84^{+2.48}_{-2.53}$& $11.43^{+0.88}_{-3.00}$      \\
QG     & power-law & ML & 3.77 & [40 , 2000] & $4.01^{+2.96}_{-2.91}$ & $5.54^{+1.92}_{-2.77}$ & $4.37^{+1.39}_{-2.01}$&$3.32^{+2.24}_{-2.31}$& $11.13^{+0.96}_{-4.01}$      \\
CCF      & DM & ML  & 3.77 & [10 , 1000] &- & - & $3.94^{+1.58}_{-2.94}$&$2.73^{+2.44}_{-2.55}$& $10.70^{+1.28}_{-10.70}$     \\    
      & DM$'$ & ML  & 3.77 & [10 , 1000] &- & - & $4.48^{+1.75}_{-3.18}$&$2.73^{+2.36}_{-2.49}$& -     \\  
      & DM & ML  & 3.77 & [40 , 160] &- & - & $7.31^{+1.86}_{-2.32}$&$9.39^{+4.86}_{-4.67}$& $12.89^{+0.54}_{-1.08}$     \\    
      & DM & ML  & 3.77 & [40 , 1000] &- & - & $4.52^{+1.46}_{-2.19}$&$3.59^{+2.47}_{-2.60}$& $11.29^{+0.94}_{-4.29}$     \\   
      & DM & ML  & 3.77 & [40 , 2000] &- & - & $4.49^{+1.44}_{-2.13}$&$3.54^{+2.42}_{-2.52}$& $11.26^{+0.95}_{-4.08}$     \\    
  \hline
  \end{tabular}}\label{tab:bias}
  \begin{tabnote}
  * The model with a prime means that it considers the contamination of the quasar and the LBG samples.
  \end{tabnote}
\end{table*}

The binned CCF is fitted through the $\chi^{2}$ minimization with a single power-law model
\begin{equation}
\omega(\theta)=A_{\omega}\theta^{-\beta}-{\rm IC}.
\end{equation}
We apply a $\beta$ of 0.86, which is determined with the ACF of the LBGs in the following section~\ref{sec:ACF}. ${\rm IC}$ is the integral constraint which is a negative offset due to the restricted area of an observation \citep{GP1977}. As described in \citet{Roche2002}, the integral constraint can be estimated by integrating the true $\omega(\theta)$ on the total survey area $\Omega$ as
\begin{equation}
{\rm IC}=\frac{1}{\Omega^{2}}\int\int\omega(\theta)d\Omega_{1}d\Omega_{2}.
\end{equation}
We calculate the integral constraint using random LBG-random LBG pairs over the entire survey area through
\begin{equation}
{\rm IC}=\frac{\sum{[RR(\theta)A_{\omega}\theta^{-\beta}]}}{\sum{RR(\theta)}}
\end{equation}
following \citet{Roche2002}. Since the survey area is wide and the scale of interest is within \timeform{1000.0''}, the ${\rm IC}/A_{\omega}$ is small compared to the observed CCFs and the IC term can be neglected in the fitting process. 

In this study, we focus on the large scale clustering between two halos, i.e. two-halo term. Thus the excess within an individual halo (one-halo term) is not considered in the fitting process. The radial scale of the region dominated by the one-halo term is examined to be $0.2-0.5$ comoving $h^{-1}$Mpc (e.g. \cite{ouchi2005}; \cite{KO2012}). At redshift 4, the corresponding angular separation is $\sim$\timeform{10.0''} - \timeform{20.0''}. Thus we fit the binned CCF with $A_{\omega}$ in the scale larger than \timeform{10.0''}. The best fit $A_{\omega}$ is summarized in table~\ref{tab:bias} where the upper and lower limits correspond to $\triangle \chi^{2}=1$ from the minimal $\chi^{2}$. Here, the $\chi^{2}$ fitting fails to fit the CCF of the SDSS luminous quasars with negative bins due to the limited luminous quasar sample size.

Another fitting method, the maximum likelihood (ML) method, which does not require a specific binning is applied to the CCFs since the $\chi^{2}$ fitting to the binned CCFs can be highly affected by the negative bins. As described in \citet{Croft1997}, if we assume that the pair counts in each bin follows the Poisson distribution, we can define a likelihood of having the observed pair sample from a model of a correlation function as
\begin{equation}
\mathcal{L}=\prod_{i=1}^{N_{bins}}\frac{e^{-h(\theta_{i})}h(\theta_{i})^{\langle DD(\theta_{i}) \rangle}}{{\langle DD(\theta_{i}) \rangle}!},
\end{equation}
where $h(\theta)=(1+\omega(\theta))\langle DR(\theta)\rangle$ is the expected object-object mean pair counts evaluated from the object-random object pair counts within a small interval around $\theta$. Here, $\omega(\theta)$ is the power-law model (equation (15)). Then, we can define a function for minimization, $S\sim-2$ln$\mathcal{L}$, as 
\begin{equation}
S=2\sum_{i}^{N_{bins}} h(\theta_{i})-2\langle DD(\theta_{i}) \rangle\sum_{i}^{N_{bins}} \ln h(\theta_{i}),
\end{equation}
where only terms dependent on model parameters are kept. Assuming that $\triangle S$ follows a $\chi^{2}$ distribution with one degree of freedom, the parameter range with $\triangle S=1$ from the minimum value corresponds to a 68\% confidence range of the parameter. 

The ML fitting is applied for the CCFs in the range between \timeform{10.0''} and \timeform{1000.0''} with an interval of \timeform{0.5''}. The interval is set to keep the object-object pair count in each bin small enough, so that the bins are independent of each other. The best fit parameters are summarized in table~\ref{tab:bias}. The ML method yields slightly higher $A_{\omega}$ than the $\chi^{2}$ fitting but still consistent within the 1$\sigma$ uncertainty. However, in the range containing several negative bins, the best ML fitting models can be lower than the positive bins of the binned CCF, as can be seen in the right panel of figure~\ref{fig:CCF}. It is reported that the assumption that pair counts follow the Poisson statistics (i.e., clustering is negligible) will underestimate the uncertainty of the fitting \citep{Croft1997}. We find the scatter of the ML fitting is only slightly smaller than the $\chi^{2}$ fitting. Therefore, we adopt the ML fitting results hereafter for both of the CCFs since both of them have negative bins in the binned CCFs.

\begin{table*}
 \tbl{HSC LBG ACF at z$\sim$4}{
  \begin{tabular}{cccccccccccc}
     \hline
     $\theta$(arcsec) &($\theta_{min}$, $\theta_{max}$) &$\langle DD \rangle$& $\langle DR \rangle$ & $\omega(\theta)$ & $\sigma(\theta)$&$\theta$(arcsec)&($\theta_{min}$, $\theta_{max}$) &$\langle DD \rangle$& $\langle DR \rangle$ & $\omega(\theta)$ & $\sigma(\theta)$  \\
\hline
2.05 & (1.58, 2.51) &16 & 25 & 6.45 & 3.70&51.45 & (39.81, 63.10) & 966 & 9376 & 0.21 & 0.05\\
3.25 &  (2.51, 3.98) &16 & 54 & 2.47 & 1.27&81.55 & (63.10, 100.00) & 2211 & 22983 & 0.12 & 0.03\\
5.15 & (3.98, 6.31) & 20 & 122 & 0.92 & 0.46&129.24 & (100.00, 158.49) & 5413 & 56115 & 0.13 & 0.02\\
8.15 &  (6.31, 10.00) &48 & 285 & 0.96 & 0.40&204.84 &  (158.49, 251.19) &12542 & 138926 & 0.06 & 0.01\\
12.92 &(10.00, 15.85) & 105 & 683 & 0.80 & 0.20&324.65 &(251.19, 398.11) & 30387 & 341510 & 0.04 & 0.01\\
20.48 & (15.85, 25.12) & 219 & 1601 & 0.60 & 0.17&514.53 & (398.11, 630.96) & 74669 & 843464 & 0.04 & 0.008\\
32.46 & (25.12, 39.81) & 410 & 3833 & 0.25 & 0.90&815.48 & (630.96, 1000.0) & 181116 & 2070430 & 0.02 & 0.007\\
\hline
 \end{tabular}}\label{tab:LBG-ACF}
\end{table*}

The contamination rates of the HSC quasar and LBG samples are taken into account by 
\begin{equation}
A'_{\omega}=\frac{A^{\rm fit}_{\omega}}{(1-f^{\rm QSO}_{c})(1-f^{\rm LBG}_{c})},
\end{equation}
where $f^{\rm QSO}_{c}$ and $f^{\rm LBG}_{c}$ are the contamination rates of the less-luminous quasar and LBG samples estimated by sections~\ref{sec:HSCz4-qso} and \ref{sec:HSCz4-lbg-z}, respectively. Since we do not know redshift distributions and clustering properties of the contaminating sources, we simply assume that they are randomly distributed in the survey area. The $A_{\omega}$ after correcting for the contamination is listed in table~\ref{tab:bias}. We note that the contaminating galaxies or galactic stars can have their own spatial distributions. For example, it is reported that the galactic stars cause measurable deviation from the true correlation function only on scales of a degree or more due to their own clustering property (e.g. \cite{Myers2006}; \cite{Myers2007}). Therefore the correction in this work only gives an upper limit of the true $A_{\omega}$ and we rely on the values without the correction in the discussions.

\subsection{Auto-correlation function of $z\sim4$ LBGs}
\label{sec:ACF}

\begin{figure}
 \begin{center}
   \includegraphics[width=0.45\textwidth]{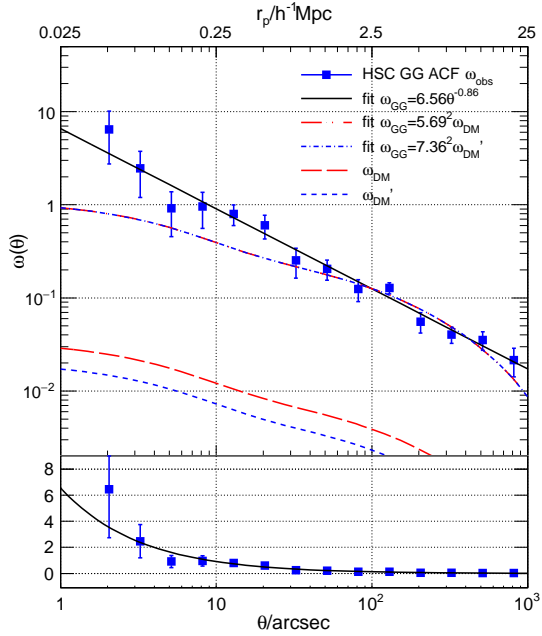}
 \end{center}
 \caption{Blue squares are the observed ACF $\omega_{\rm obs}$ of the LBGs at $z\sim4$. Solid line is the best fit power-law model in the scale \timeform{10.0''} to \timeform{1000.0''}. Red dash-dotted line is the best fit dark matter model $\omega_{\rm DM}$ (red long-dashed line) in the same scale based on the HALOFIT power spectrum \citep{smith2003} following the method in \citet{Myers2007}, while the blue dash-dotted line is the best fit dark matter model $\omega_{\rm DM}'$ (blue short-dashed line) after considering the contamination of the LBGs. The $\chi^{2}$ fitting results are shown. Top and bottom panels show the logarithmic and the linear scale of the vertical axis respectively. Top horizontal axis of the top panel is the comoving distance at redshift 4. 
}\label{fig:LBG-ACF}
\end{figure}

In order to derive the bias factor of the quasars from the strength of the quasar-LBG CCFs, we need to evaluate the bias factor of the LBGs from the LBG ACF. The binned ACF of the $z\sim4$ LBGs is derived in the same way as the quasar-LBG CCF. We use the estimator
\begin{equation}
\omega(\theta)=\frac{DD(\theta)}{DR(\theta)}-1,
\end{equation}
where $DD(\theta)=\langle DD \rangle/(N_{LBG}(N_{LBG}-1)/2)$ and $DR(\theta)=\langle DR \rangle/N_{LBG}N_{R}$ are the normalized LBG-LBG and LBG-random LBG pair counts in the annulus between $\theta-\Delta \theta$ and $\theta+\Delta \theta$, respectively. Here, $\langle DD \rangle$ and $\langle DR \rangle$ are the numbers of LBG-LBG and LBG-random LBG pairs in the annulus, and $N_{LBG}$ and $N_{R}$ are the total numbers of LBGs and random LBGs, respectively. We set 14 bins from \timeform{1.0"} to \timeform{1000.0"} in the logarithmic scale. The LBG ACF is shown in figure~\ref{fig:LBG-ACF} and table~\ref{tab:LBG-ACF} along with the pair counts. Thanks to the large sample of the LBGs, the LBG-LBG pair count is large enough to constrain the ACF even in the smallest bin. We adopt the Jackknife error, which has two times larger value than the Poisson error at all bins. Most of the bins have clustering signal more than 3$\sigma$.  

\begin{figure}
 \begin{center}
   \includegraphics[width=0.45\textwidth]{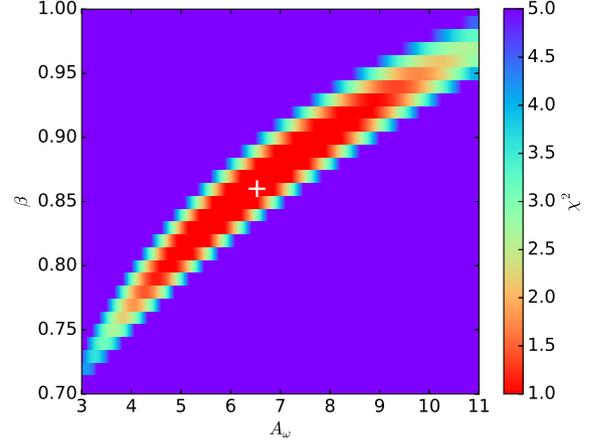}
 \end{center}
\caption{$\chi^{2}$ map of $A_{\omega}$ and $\beta$ parameter of the ACF of the LBGs. White cross indicates the best fit $A_{\omega}$ and $\beta$ at the minimal $\chi^{2}$, while the red region indicates the 68\% confidence region.
 }\label{fig:beta}
\end{figure}

We fit the raw LBG ACF with a single power-law model
$\omega(\theta)=A_{\omega}\theta^{-\beta}-{\rm IC}$ by $\chi^{2}$ minimization in the scale from \timeform{10.0''} to \timeform{1000.0''}. The integral constraint is negligible. Thanks to the small uncertainty of the LBG ACF, the power-law index can be constrained tightly to be $\beta=0.86^{+0.07}_{-0.06}$ as shown in figure~\ref{fig:beta}. As already mentioned in section 3.1, we adopt this power-law index throughout this paper. The best fit parameters are listed in table~\ref{tab:bias-lbg}. 

The effect of the contamination is evaluated with
\begin{equation}
A'_{\omega}=\frac{A^{\rm fit}_{\omega}}{(1-f^{\rm LBG}_{c})^{2}}.
\end{equation}
The results are listed in table~\ref{tab:bias-lbg}. We do not consider the contamination for fitting the power-law index $\beta$ because it would not be affected by a random contamination.

\begin{table*}
 \tbl{Summary of the clustering analysis of HSC LBGs ACF}{
 \begin{tabular}{ccccccccc}
 \hline
& model&fitting& $\bar{z}$ & [$\theta_{min},\theta_{max}$] &$\beta$& $A_{\omega}$ & $r_{0}$       &  bias \\
 &            &       &           & (arcsec)           &          &             &($h^{-1}$ Mpc) &       \\
 \hline
 &power-law & $\chi^{2}$ & 3.71 & [10 , 1000] &$0.86^{+0.07}_{-0.06}$ &$6.56^{+0.49}_{-0.49}$ & $7.47^{+0.29}_{-0.31}$& $5.76^{+0.21}_{-0.22}$ \\
LBG &power-law$'$ $*$& $\chi^{2}$ & 3.71 & [10 , 1000] &$0.86^{+0.07}_{-0.06}$ &$10.97^{+0.82}_{-0.82}$ & $9.85^{+0.39}_{-0.40}$& $7.45^{+0.27}_{-0.28}$ \\
ACF & DM & $\chi^{2}$  & 3.71 & [10 , 1000] &-& -& - & $5.69^{+0.21}_{-0.22}$\\
 & DM$'$& $\chi^{2}$  & 3.71 & [10 , 1000] &-& -& - & $7.36^{+0.27}_{-0.28}$\\
    \hline
  \end{tabular}}\label{tab:bias-lbg}
  \begin{tabnote}
  * The model with a prime means that it considers the contamination of the LBG sample.
  \end{tabnote}
\end{table*}

\section{Discussion}
\label{sec:discussion}

\subsection{Clustering bias from the correlation length}
\label{sec:bias}

One of the parameters representing the clustering strength is the spatial correlation length, $r_{0}$ ($h^{-1}$ Mpc), which is in the spatial correlation function with the power-law form as 
\begin{equation}
\xi(r)=(\frac{r}{r_{0}})^{-\gamma},
\end{equation}
where $\gamma$ is related to the power of the projected correlation function through $\gamma=1+\beta$. The spatial correlation function can be projected to the angular correlation function through Limber's equation (\cite{limber1953}). We ignore the redshift evolution of the clustering strength within the covered redshift range. Then the spatial correlation length of the ACF can be derived from the amplitude of the angular correlation function, $A_{\omega}$, as
\begin{equation}
r_{0}=[ A_{\omega} \frac{c}{H_{0}H_{\gamma}} \frac{[\int N(z)dz]^{2}}{\int N^{2}(z)\chi(z)^{1-\gamma}E(z)dz} ]^{1/\gamma},
\end{equation}
where 
\begin{equation}
H_{\gamma}=\frac{\Gamma(\frac{1}{2})\Gamma(\frac{\gamma-1}{2})}{\Gamma(\frac{\gamma}{2})},
\end{equation}
\begin{equation}
E(z)=[\Omega_{m}(1+z)^{3}+\Omega_{\Lambda}]^{1/2},
\end{equation}
\begin{equation}
\chi(z)=\frac{c}{H_{0}}\int_{0}^{z}\frac{1}{E(z')}dz'
\end{equation}
and $N(z)$ is the redshift distribution of the sample. For the CCF, the same relation can be modified to \citep{CS1999}
\begin{equation}
r_{0}=[ A_{\omega} \frac{c}{H_{0}H_{\gamma}} \frac{\int N_{\rm QSO}(z)dz\int N_{\rm LBG}(z)dz}{\int N_{\rm QSO}(z)N_{\rm LBG}(z)\chi(z)^{1-\gamma}E(z)dz} ]^{1/\gamma}.
\end{equation}
Applying the redshift distributions of the less-luminous quasars, the luminous quasars and the LBGs at $z\sim4$ estimated in section~\ref{sec:HSCz4-qso} for $N_{\rm QSO}(z)$ and section~\ref{sec:HSCz4-lbg-z} for $N_{\rm LBG}(z)$, respectively, we evaluate $r_{0}$ from $A_{\omega}$ with and without the contamination correction as summarized in table~\ref{tab:bias}. Although the contamination rates of the less-luminous quasars and the LBGs are not high, the correlation lengths of the less-luminous quasar-LBG CCF and the LBG ACF are significantly increased after correcting for the contamination. Meanwhile, $r_{0}$ of the luminous quasar-LBG CCF vary slightly, because the SDSS quasar sample is not affected by a contamination. 

The measurement of $r_{0}$ is sensitive to the assumed redshift distribution of the sample. For example, $r_{0}$ will be smaller if we assume a narrower redshift distribution even for the same $A_{\omega}$. As discussed in section~\ref{sec:HSCz4-lbg-z}, the redshift distribution of the LBGs is estimated to be more extended than both of the less-luminous and luminous quasar samples. If we assume the redshift distribution of the LBGs is the same as the less-luminous quasars, $r_{0}$ of the LBG and the less-luminous quasars decreases to $5.52^{+0.77}_{-0.87}$ $h^{-1}$Mpc, which is 23\% lower than that estimated originally, because the fraction of the LBGs contributing to the projected correlation function in the overlapped redshift range increases, yielding a weaker correlation strength, i.e. a smaller $r_{0}$ from a fixed $A_{\omega}$.

The bias factor is defined as the ratio of clustering strength of real objects to that of the underlying dark matter at the scale of 8 $h^{-1}$Mpc,
\begin{equation}
b=\sqrt{\frac{\xi(8,z)}{\xi_{DM}(8,z)}}.
\end{equation}
The clustering strength of the underlying dark matter can be evaluated based on the linear structure formation theory under the cold dark matter model \citep{Myers2006} as 
\begin{equation}
\xi_{DM}(8,z)=\frac{(3-\gamma)(4-\gamma)(6-\gamma)2^{\gamma}}{72}[\sigma_{8}\frac{g(z)}{g(0)}\frac{1}{z+1}]^{2},
\end{equation}
where 
\begin{equation}
g(z)=\frac{5\Omega_{mz}}{2}\Bigl [\Omega^{4/7}_{mz}-\Omega_{\Lambda z}+(1+\frac{\Omega_{mz}}{2})(1+\frac{\Omega_{\Lambda z}}{70})\Bigl ]^{-1},
\end{equation}
and
\begin{equation}
\Omega_{mz}=\frac{\Omega_{m}(1+z)^{3}}{E(z)^{2}},\Omega_{\Lambda z}=\frac{\Omega_{\Lambda}}{E(z)^{2}}.
\end{equation}
We derive the bias factors $b_{\rm LBG}$ and $b_{\rm QG}$ from the spatial correlation length of the LBG ACF and the quasar-LBG CCF, respectively. Following \citet{Mountrichas2009}, the quasar bias factor is then evaluated from the bias factor of the CCF by
\begin{equation}
b_{\rm QSO}b_{\rm LBG}\sim b^{2}_{\rm QG}.
\end{equation}
We list the LBG ACF bias factors in table~\ref{tab:bias-lbg}.The estimated $b_{\rm LBG}$ with and without the contamination correction are consistent with \citet{Allen2005} and the brightest bin at $M_{UV}\sim-21.3$ in \citet{ouchi2004}, respectively. The quasar bias factors derived from the CCF are summarized in table~\ref{tab:bias}. 

\subsection{Bias factor from comparing with the HALOFIT power spectrum}
\label{sec:DM}

The bias factors can also be derived by directly comparing the observed clustering with the predicted clustering of the underlying dark matter from the power spectrum $\Delta^{2}(k,z)$ (e.g., \cite{Myers2007}). The spatial correlation function derived from $\Delta^{2}(k,z)$ can be projected with the Limber's equation into the angular correlation $\omega_{\rm DM}(\theta)$ as  
\begin{equation}
\omega_{\rm DM}(\theta)=\pi\int \int \frac{\Delta^{2}(k,z)}{k}J_{0}[k\theta\chi(z)]N^{2}(z)\frac{dz}{d\chi}F(\chi)\frac{dk}{k}dz,
\label{myers}
\end{equation}
where $J_{0}$ is the zeroth-order Bessel function, $\chi$ is the radial comoving distance, $N(z)$ is the normalized redshift distribution function, $dz/d\chi=H_{z}/c=H_{0}[\Omega_{m}(1+z)^{3}+\Omega_{\Lambda}]^{1/2}/c$, and $F(\chi)=1$ for the flat universe. We evaluate the non-linear evolution of the power spectrum $\Delta_{NL}^{2}(k, z)$ in the redshift range between $z=3$ and $5$ with the HALOFIT code \citep{smith2003} by adopting the cosmological parameters used throughout this paper. The bias parameters are derived by fitting $b^{2}\omega_{\rm DM}(\theta)$ to the observed correlation functions, $\omega_{\rm obs}(\theta)$. For the LBG ACF, $\omega_{\rm DM}(\theta)$ is directly compared to the $\omega_{\rm obs}(\theta)$ through $\chi^{2}$ minimization. For the CCFs, the redshift distribution in equation \ref{myers} is replaced by the multiplication of those of quasars and LBGs as
\begin{eqnarray}
\omega_{\rm DM-CCF}(\theta)=\pi\int \int & &\frac{\Delta^{2}(k,z)}{k}J_{0}[k\theta\chi(z)]N_{\rm QSO}(z)N_{\rm LBG}(z) \nonumber \\
& &\frac{dz}{d\chi}F(\chi)\frac{dk}{k}dz.
\end{eqnarray}
In the scale from \timeform{10.0''} to \timeform{1000.0''}, both of the $\chi^{2}$ and ML fitting are applied to the less-luminous quasar CCF, while only ML fitting works for the luminous quasar CCF. The bias factors of the quasar samples are derived from the CCF and the LBG ACF through equation (33). The best fit bias factors are summarized in table~\ref{tab:bias} and table~\ref{tab:bias-lbg}. They are consistent with those derived from the power-law fitting within the $1\sigma$ uncertainty. Thus the power law approximation with an index of $\beta=-0.86$ can well reproduce the underlying dark matter distribution on scales larger than \timeform{10.0''}. 

In the scale below \timeform{10.0''}, the underlying dark matter model becomes flat since we do not consider the one-halo term. If we compare the observed correlation functions with the best-fit power-spectrum models, there is an obvious overdensity of galaxies in that scale in figure~\ref{fig:LBG-ACF}, which is consistent with the one-halo term of the LBG ACF at $z\sim4$ (e.g., \cite{ouchi2005}). In the left panel of figure~\ref{fig:CCF}, it also shows an overdensity of galaxies within \timeform{10.0''} around the less-luminous quasars although the error bar is large. Interestingly, we find that the luminous quasars show no pair count within \timeform{10.0''} in the right panel of figure~\ref{fig:CCF}. It should be noted that the best fit model in scales larger than \timeform{10.0''} suggests only 1 SDSS quasar - HSC LBG pair within \timeform{10.0''}, which is consistent with no pair count. Thus it can be caused by the limited size of the SDSS quasar sample, though we cannot exclude the possibility that there is a real deficit of galaxies around luminous quasars within \timeform{10.0''}. 

We consider the contamination by modifying the redshift distribution normalization $\int_{0}^{\infty}N(z)dz\sim1-f_{c}$ for the less-luminous quasars and the LBGs respectively. We simply assume that the contamination will not contribute to the underlying dark matter correlation function. The modified underlying dark matter correlation functions are plotted in figures~\ref{fig:CCF} and \ref{fig:LBG-ACF}. Since the redshift distribution form is the same after considering the contamination, only the amplitude of the underlying dark matter correlation function is changed. The bias factors with contamination are listed in table~\ref{tab:bias} and table~\ref{tab:bias-lbg}, which are consistent with those derived from fitting with the power-law model after correcting for the contamination. 

\subsection{Redshift and luminosity dependence of the bias factor}
\label{sec:luminosity}

At first, we discuss the luminosity dependence of the bias factors of the luminous and less-luminous quasars in this work. The bias factor of the less-luminous quasars is $5.93^{+1.34}_{-1.43}$, which is derived by fitting the CCF with the underlying dark matter model in the scale from \timeform{10.0''} to \timeform{1000.0''} through the ML fitting. The bias factor is consistent with that of the luminous quasars, $2.73^{+2.44}_{-2.55}$, obtained from the CCF through the same method within the 1 $\sigma$ uncertainty. If we consider the possible effect of the contamination, the bias factor of the less-luminous quasars increases to $6.58^{+1.49}_{-1.58}$, which is still consistent with that of the luminous quasars within the uncertainty. Thus no or only a weak luminosity dependence of the quasar clustering is detected within the two samples.

\begin{figure*}
 \begin{center}
   \includegraphics[width=0.45\textwidth]{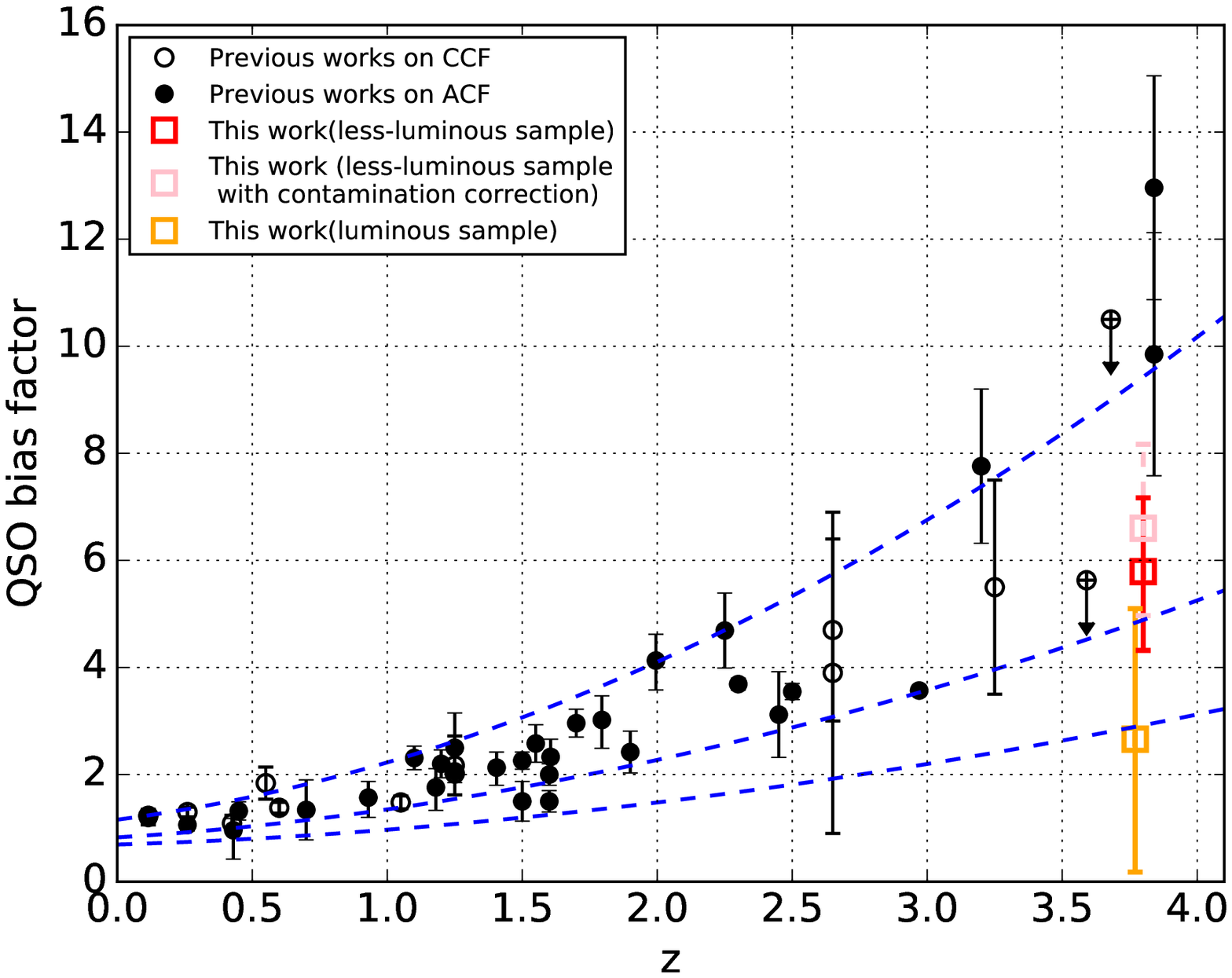}
   \includegraphics[width=0.45\textwidth]{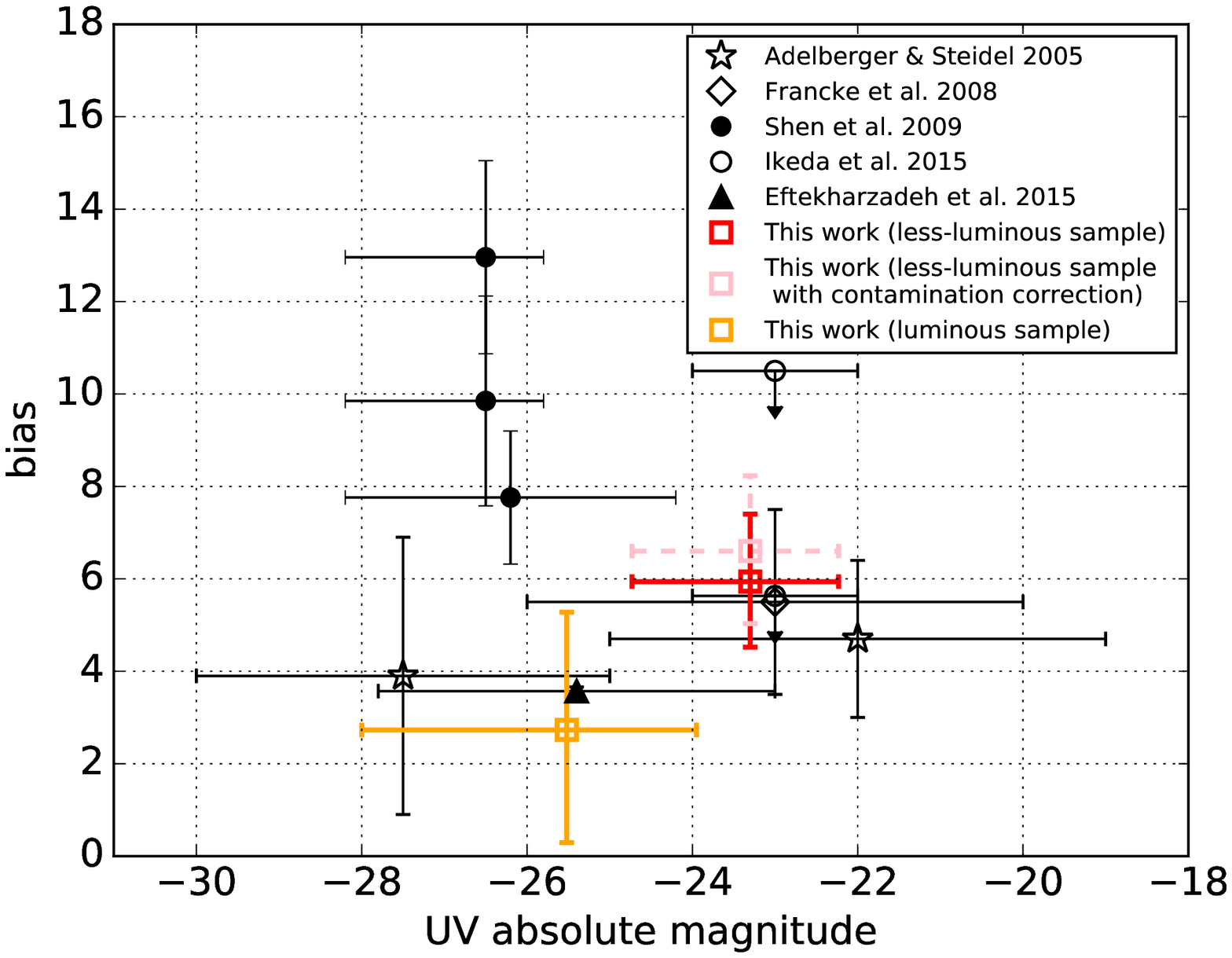}
 \end{center}
\caption{Left panel: the redshift evolution of the quasar bias factor. Red square is the one from fitting the less-luminous quasar CCF against the underlying dark matter model through ML fitting. Pink square is derived from fitting the less-luminous quasar CCF with the same method after considering the contamination. Orange square is obtained from fitting the luminous quasar CCF with the same method. Open and filled black circles are bias factors of quasars in a wide luminosity range obtained from the CCF and the ACF, respectively, in the literature, which are summarized by \citet{ikeda2015} and \citet{Eftekharzadeh2015}. Blue dashed lines show the bias evolutions of halos with a fixed mass of $10^{11}$, $10^{12}$, $10^{13}h^{-1}M_{\odot}$ from bottom to top following the fitting formulae in \citet{sheth2001}. Right panel: the luminosity dependence of the quasar bias at $3<z<5$. Red and orange squares have the same meaning with the left panel. The stars, diamonds, dots, triangle, open circles and squares are from \citet{AS2005},  \citet{Francke2007}, \citet{shen2009}, \citet{Eftekharzadeh2015}, \citet{ikeda2015} and this work. Open and filled symbols imply the bias factor is derived from the CCF and the ACF, respectively. 
 }\label{fig:luminosity-bias}
\end{figure*}

In order to discuss the redshift dependence of the quasar clustering, we compare the bias factors with those in the literature in the left panel of figure~\ref{fig:luminosity-bias}.
The bias factors in the previous studies show a trend that quasars at higher redshifts are more strongly biased, indicating that quasars preferentially reside in DMHs within a mass range of $10^{12}\sim10^{13}h^{-1}M_{\odot}$ from $z\sim0$ to $z\sim4$. There is no discrepancy between the bias factors estimated with the ACF and the CCF at $z\lesssim3$. In this work, the bias factor of the less-luminous quasars at $z\sim4$ follows the trend, while the bias factor of the luminous quasars is similar to or even smaller than those at $z\sim3$.

The luminosity dependence of the quasar bias factors at $z\sim3-4$ is summarized in the right panel of figure~\ref{fig:luminosity-bias}. Both of the bias factors of the less-luminous quasars with and without the contamination correction are consistent with but slightly higher than that evaluated with the CCF of 54 faint quasars in the magnitude range of $-25.0<M_{UV}<-19.0$ at $1.6<z<3.7$ measured by \citet{AS2005}, the CCF of 58 faint quasars in the magnitude range of $-26.0<M_{UV}<-20.0$ at $2.8<z<3.8$ measured by \citet{Francke2007}, and the CCF of 25 faint quasars in the magnitude range of $-24.0<M_{UV}<-22.0$ at $3.1<z<4.5$ measured by \citet{ikeda2015}, which suggests a slightly increasing or no evolution from $z=3$ to $z=4$. 

Meanwhile, for the clustering of the luminous quasars, the bias factor in this work is consistent with the CCF of 25 bright quasars in the magnitude range of $-30.0<M_{UV}<-25.0$ at $1.6<z<3.7$ measured by \citet{AS2005} and the ACF of 24724 bright quasars in the magnitude range of $-27.81<M_{UV}<-22.9$ mag at $2.64<z<3.4$ measured by \citet{Eftekharzadeh2015}. Different from the case of the less-luminous quasars, the clustering of the luminous quasars suggests no or a declining evolution from $z\sim3$ to $z\sim4$. The bias factor of the luminous quasars in this work shows a large discrepancy with the ACF of 1788 bright quasars in the magnitude range of $-28.2<M_{UV}<-25.8$ (which is transferred from $M_{i}(z=2)$ by equation (3) in \citet{richards2006}) at $3.5<z<5.0$ measured by \citet{shen2009}. They give two values for the bias factor that the higher one is obtained by only considering the positive bins and the lower one considers all of the bins in the ACF. The bias factor from another subsample of bright quasars covering $-28.0<M_{UV}<-23.95$ at $2.9<z<3.5$ in \citet{shen2009} is also shown in the panel. 

\begin{figure}
 \begin{center}
   \includegraphics[width=0.45\textwidth]{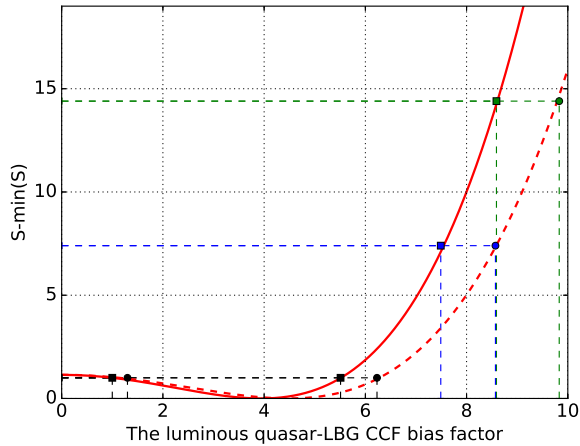}
 \end{center}
\caption{ML fitting minimization function $S$ of fitting for the luminous quasar CCF with the dark matter model. $S$ is shown in relative to the minimal value. Black squares mark the 68\% upper and lower limits of $b_{QG}$ with $S-{\rm min}(S)=1$. Blue and green squares indicate the expected bias factors of $b_{QG}\sim\sqrt{b_{QSO}b_{LBG}}=7.53$ and $8.64$ from the bias factors of the SDSS luminous quasars in \citet{shen2009} with and without considering the negative bins in the ACF, respectively. The red dashed line indicates the same minimization function $S$ after considering the possible contamination in the LBG sample. Black, blue and green dots have the same meaning with the squares.
 }\label{fig:pdf}
\end{figure}

The $z\sim4$ quasar bias factors in \citet{shen2009} show a large discrepancy from the bias factor of the luminous quasars in this work and in \citet{Eftekharzadeh2015} with the similar magnitude and redshift coverage. In the right panel of figure~\ref{fig:CCF}, we plot the expected CCF with $b_{\rm QG}\sim\sqrt{b_{\rm QSO}b_{\rm LBG}}=9.83$ by the orange dash-dot-dotted line. We adopt the higher $b_{\rm QSO}$ in \citet{shen2009} and the $b_{\rm LBG}$ with the contamination correction to measure the upper limit of the $b_{\rm QG}$. Although the expected CCF is consistent with some bins within the 1$\sigma$ uncertainty, it predicts much stronger clustering than both of the best fit power-law and dark matter models. In order to quantitatively examine the discrepancy, we plot the minimization function $S$ of the ML fitting for the luminous quasars with the HALOFIT power spectrum as a function of the bias factor in figure~\ref{fig:pdf}. Both of the bias factors at $3.5<z<5$ in \citet{shen2009} are beyond the 1$\sigma$ uncertainty, corresponding to a low probability. 
Meanwhile, the bias factor in \citet{Eftekharzadeh2015}, whose uncertainty is small thanks to the large sample, also shows a large discrepancy from those in \citet{shen2009}. \citet{Eftekharzadeh2015} suspect the discrepancy is mainly caused by a difference in large scale bins ($>30h^{-1}$Mpc). We further investigate the effect from the fitting scale as shown in table~\ref{tab:CCF} and the right panel of figure~\ref{fig:CCF}. In the scale of \timeform{40.0''} to \timeform{160.0''}, we find a strong CCF of the luminous quasars and the LBGs, which is consistent with the ACF of the luminous quasars. On scales below \timeform{40.0''}, the ML fitting suggests a $b_{\rm QG}$ of $0$. On larger scales, the ML fitting is not efficient since the pair counts in each bin is too large to fulfill the assumption that bins are independent with each other, even if choosing a small bin width of \timeform{0.5''} interval. Therefore we only expand the ML fitting scale to \timeform{2000.0''}. If we consider the power-law model, the $b_{\rm QG}$ obtained by fitting in the range of \timeform{40.0''} to \timeform{1000.0''} is 24.7\% and 7.6\% higher than that estimated in the range of \timeform{10.0''} to \timeform{1000.0''} and of \timeform{40.0''} to \timeform{2000.0''}, respectively, which suggests that the deficit of the luminous quasar-LBG pair on small scales may weaken the CCF more severely than fitting on scales larger than \timeform{1000.0''}. Such deficit can be an implication of the feedback from the luminous quasars. Since the fitting of the luminous quasar CCF strongly depends on the scale, especially on small scales, we still focus on the results in the scale of \timeform{10.0''} to \timeform{1000.0''} to keep accordant to the LBG ACF and the less-luminous quasar CCF throughout the discussion.

Quasar clustering models based on semi-analytic galaxy models predict no luminosity dependence of the quasar clustering at redshift 4 (e.g. \cite{Fanidakis2013}; \cite{oogi2016}). Although there is a relation between mass of the SMBHs and DMHs in the models, SMBHs in a wide mass range are contributing to quasars at a fixed luminosity, thus there is no relation between the luminosity of model quasars and the mass of their DMHs. The predicted quasar bias factor at redshift 4 in \citet{oogi2016} is $3.0\sim5.0$, which is consistent with the quasar bias factors in this work. No luminosity dependence is also predicted in a continuous SMBH growth model of \citet{hopkins2007}. They assume an Eddington limited SMBH growth until redshift 2. However, the predicted bias factor is much larger than the results in this work. On the other hand, there are models which predict stronger luminosity dependence of the quasar clustering at higher redshifts (e.g. \cite{shen2009b}; \cite{CW2012}). These models predict SMBHs in a narrow mass range are contributing to the luminous quasars.

In order to disclose the luminosity and redshift dependences of the quasar clustering, we need to understand the cause of the discrepancy between the quasar ACF and quasar-LBG CCF for the luminous quasars at $z\sim4$. The quasar-LBG CCF could be affected by the suppression of galaxy formation due to feedback from luminous quasars (e.g. \cite{kashikawa2007}; \cite{Utsumi2010}; \cite{Uchiyama2017}). The weak cross-correlation could also be induced by a discrepancy between the redshift distributions of the quasars and LBGs. We need to further determine the redshift distribution through spectroscopic follow-up observations of the LBGs.

\subsection{DMH mass}
\label{sec:DMHmass}

The bias factor of a population of objects is directly related to the typical mass of their host DMHs, because more massive DMHs are more strongly clustered and biased in the structure formation under the $\Lambda$CDM model \citep{ST1999}. The relation between the $M_{\rm DMH}$ and the bias factor is derived based on an ellipsoidal collapse model that is calibrated by an N-body simulation as
\begin{eqnarray}
b(M,z)&=&1+\frac{1}{\sqrt{a}\delta_{crit}}[a\nu^{2}\sqrt{a}+b\sqrt{a}(a\nu^{2})^{(1-c)} \nonumber \\
& &-\frac{(a\nu^{2})^{c}}{(a\nu^{2})^{c}+b(1-c)(1-c/2)}],
\end{eqnarray}
where $\nu=\delta_{crit}/(\sigma(M)D(z))$ and critical density $\delta_{crit}=1.686$ \citep{sheth2001}. We adopt the updated parameters $a=0.707, b=0.35, c=0.80$ in \citet{Tinker2005}. The rms mass fluctuation $\sigma(M)$ on a mass scale $M$ at redshift 0 is given by
\begin{equation}
\sigma^{2}(M)=\int\Delta^{2}(k)\tilde{W}^{2}(kR)\frac{dk}{k},
\end{equation}
and
\begin{equation}
M(R)=\frac{4\pi \overline{\rho_{0}} R^{3}}{3},
\end{equation}
where $R$ is the comoving radius, $\tilde{W}(kR)=(3\sin(kR)-(kR)\cos(kR))/(kr)^{3}$ is the top hat window function in Fourier form and $\overline{\rho_{0}}=2.78\times10^{11}\Omega_{m}h^{2}  M_{\odot}$ Mpc$^{-3}$ is the mean density in the current universe. The linear power spectrum $\Delta^{2}(k)$ at redshift 0 is obtained 
from the HALOFIT code \citep{smith2003}. The growth factor $D(z)$ is approximated by 
\begin{equation}
D(z)\propto\frac{g(z)}{1+z}
\end{equation}
following \citet{Carroll1992}. 

Assuming the quasars and LBGs are associated with DMHs in a narrow mass range, we can infer the mass of the quasar host DMHs through the above relations. The evaluated halo masses of the less-luminous quasars and the luminous quasars are $1\sim2\times10^{12}h^{-1}M_{\odot}$ and $<10^{12}h^{-1}M_{\odot}$ as summarized in table~\ref{tab:bias}, respectively. Since the bias factor of the luminous quasars has a large uncertainty, we could only set an upper limit of the $M_{\rm DMH}$. We note that the halo mass strongly depends on the amplitude of the power spectrum on the scale of 8 $h^{-1}$ Mpc, $\sigma_{8}$. If we adopt $\sigma_{8}=0.9$, the host DMH mass of the less-luminous quasars will be $4-6\times10^{12}h^{-1}M_{\odot}$ with the same bias factor. 

\subsection{Minimum halo mass and duty cycle}
\label{sec:duty}

In the above discussion, we assume that quasars are associated with DMHs in a specific mass range, but it may be more physical to assume that quasars are associated with DMHs with a mass above a critical mass, $M_{\rm min}$. In this case, the effective bias for a population of objects which are randomly associated with DMHs above $M_{\rm min}$ can be expressed with
\begin{equation}
b_{\rm eff}=\frac{\int_{M_{\rm min}}^{\infty}b(M)n(M)dM}{\int_{M_{\rm min}}^{\infty}n(M)dM},
\end{equation}
where $n(M)$ is the mass function of DMHs and $b(M, z)$ is the bias factor of DMHs with mass $M$ at $z$. We adopt the DMH mass function from the modified Press-Schechter theory \citep{ST1999} as 
\begin{eqnarray}
n(M,z)&=&-A\sqrt{\frac{2a}{\pi}}\frac{\rho_0}{M}\frac{\delta_c(z)}{\sigma^2(M)}\frac{d\sigma(M)}
{dM} \nonumber \\
& &\left\{1+\left[\frac{\sigma^2(M)}{a\delta_c^2(z)}\right]^p\right\}\exp[-\frac{a\delta_c^2(z)}{2\sigma^2(M)}],
\end{eqnarray}
where $A=0.3222$, $a=0.707$, $p=0.3$ and $\delta_c(z)=\delta_{crit}/D(z)$. If we follow the above formulation, the $M_{\rm min}$ is estimated to be $\sim0.3-2\times10^{12}h^{-1}M_{\odot}$ and $<5.62\times10^{11}h^{-1}M_{\odot}$ with the bias factors of the less-luminous quasars and the luminous quasars, respectively.

Comparing the number density of the DMHs above the $M_{\rm min}$ and that of the less-luminous and luminous quasars, we can infer the duty-cycle of the quasar activity among the DMHs in the mass range by
\begin{equation}
f=\frac{n_{QSO}}{\int_{M_{min}}^{\infty}n(M)dM},
\end{equation}
assuming one DMH contains one SMBH. The co-moving number density of $z\sim4$ less-luminous quasars are
estimated with the HSC quasar sample \citep{akiyama2017}. Integrating the best-fit luminosity function of $z\sim4$ quasars from $M_{\rm 1450}\sim-24.73$ to $M_{\rm 1450}\sim-22.23$, we estimate the total number density of the less-luminous quasar to be $1.07\times10^{-6}h^{3}$Mpc$^{-3}$, which is 2 times higher than that of the luminous quasars with $-28.00<M_{\rm 1450}<-23.95$ ($4.21\times10^{-7}h^{3}$Mpc$^{-3}$). If we adopt the $n(M)$ in equation (41), the duty-cycle is estimated to be $0.001\sim0.06$ and $<8\times10^{-4}$ for $M_{\rm min}$ from the less-luminous and the luminous quasar CCF, respectively. If we use the bias factor estimated by considering the effect of the possible contamination, the duty cycle of the less-luminous quasars is estimated to be $0.003\sim0.175$, which is higher than the estimation above. 

\begin{figure}
 \begin{center}
   \includegraphics[width=0.45\textwidth]{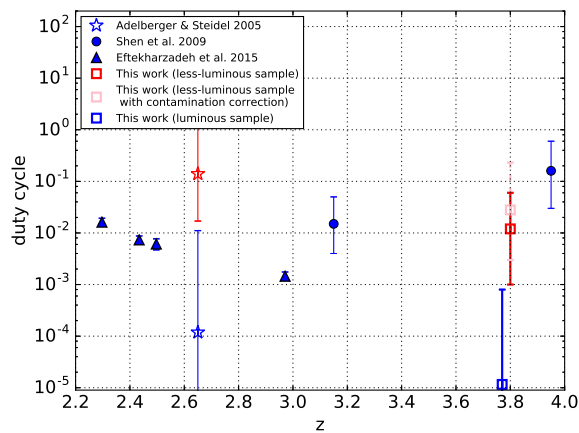}
 \end{center}
\caption{Estimated quasar duty cycle as a function of redshift. The blue symbols represent the duty cycles estimated with a sample of quasars mostly with $M_{UV}<-25$. The red symbols show those for the less-luminous quasars having $M_{UV}>-25$. Stars, triangles, filled circles and squares represent the results from \citet{AS2005}, \citet{shen2007}, \citet{Eftekharzadeh2015} and this work. The pink open square shows the duty cycle with the contamination correction.
 }\label{fig:duty}
\end{figure}

We compare the duty-cycles with those evaluated for quasars at $2<z<4$ in the literature in figure~\ref{fig:duty}. The estimated luminosity dependence of the duty-cycles is similar to that estimated for quasars in the similar luminosity range at $z\sim2.6$ \citep{AS2005}, although the duty-cycles at $z\sim4$ are one order of magnitude smaller than those at $z\sim2.6$.

The estimated duty-cycle corresponds to a duration of the less-luminous quasar activity of $1.5\sim90.8$ Myr, which is broadly consistent with the quasar lifetime range of $1\sim100$ Myr estimated in previous studies (for review see \cite{martini2004}). It needs to be noted that the estimated duty-cycle is sensitive to the measured strength of the quasar clustering. Small variation in the bias factor can results in even one order of magnitude difference in the duty-cycle, because of the non-linear relation between $b$ and $M_{\rm DMH}$ and the sharp cut-off of $n(M)$ at the high-mass end. Furthermore, the duty-cycle is also sensitive to the assumed value of $\sigma_{8}$ \citep{shen2007}.

\section{Summary}
\label{sec:summary}

We examine the clustering of a sample of 901 less-luminous quasars with $-24.73<M_{\rm 1450}<-22.23$ at $3.1<z<4.6$ selected from the HSC S16A Wide2 catalog and of a sample of 342 luminous quasars with $-28.00<M_{\rm 1450}<-23.95$ at $3.4<z_{\rm spec}<4.6$ within the HSC S16A Wide2 coverage from the 12th data release of SDSS. We investigate the quasar clustering through the CCF between the quasars and a sample of 25790 bright LBGs with $M_{\rm 1450}<-21.25$ in the same redshift range from the HSC S16A Wide2 data release. The main results are as follows.

1. The bias factor of the less-luminous quasar is $5.93^{+1.34}_{-1.43}$ derived by fitting the CCF with the dark matter power-spectrum model through the ML method, while that of the luminous quasars is $2.73^{+2.44}_{-2.55}$ obtained in the same manner. If we consider the contamination rates of 22.7\% and 10.0\% estimated for the LBG and the less-luminous quasar samples, respectively, the bias factor of the less-luminous quasars can increase to $6.58^{+1.49}_{-1.58}$ in an assumption that the contaminating objects are distributed randomly. 

2. The CCFs of the luminous and less-luminous quasars do not show significant luminosity dependence of the quasar clustering. The bias factor of the less-luminous quasars suggests that the environment around them is similar to the luminous LBGs used in this study. The luminous quasars do not show strong association with the luminous LBGs in scale \timeform{10.0''} to \timeform{1000.0''}, especially on scales smaller than \timeform{40.0''}. The bias factor of the luminous quasar is smaller than that derived from the ACF of the SDSS quasars at $z\sim4$ \citep{shen2009}. The reason may be partly due to the deficit of the pairs on small scales, which may be a reflection of the strong feedback from the SMBH. 

3. The bias factor of the less-luminous quasars corresponds to a mass of DMHs of $\sim1-2\times10^{12}h^{-1}M_{\odot}$. Minimal host DMH mass for the quasars can be also inferred from the bias factor. Combining the halo number density above that mass threshold and the observed quasar number density, the fraction of halos which are in the less-luminous quasar phase is estimated to be $0.001\sim0.06$ from the CCF. The corresponding quasar lifetime is $1.5\sim90.8$ Myr. 

Correlation analysis in this work is conducted in the projected plane, and accurate information on the redshift distribution of the samples and the contamination rates is necessary to obtain reliable constraints on the clustering of the $z\sim4$ quasars. Spectroscopic follow-up observations are expected to obtain the accurate information. Additionally, the full HSC Wide survey plans to cover 1400 deg$^2$ in 5 years, which can significantly enhance the sample size. The statistical significance of the current results can then be largely improved.

\bigskip
\begin{ack}
We would like to thank Dr. A.K. Inoue who kindly provides us with the IGM model data. 

The Hyper Suprime-Cam (HSC) collaboration includes the astronomical communities of Japan and Taiwan, and Princeton University. The HSC instrumentation and software were developed by the National Astronomical Observatory of Japan (NAOJ), the Kavli Institute for the Physics and Mathematics of the Universe (Kavli IPMU), the University of Tokyo, the High Energy Accelerator Research Organization (KEK), the Academia Sinica Institute for Astronomy and Astrophysics in Taiwan (ASIAA), and Princeton University. Funding was contributed by the FIRST program from Japanese Cabinet Office, the Ministry of Education, Culture, Sports, Science and Technology (MEXT), the Japan Society for the Promotion of Science (JSPS), Japan Science and Technology Agency (JST), the Toray Science Foundation, NAOJ, Kavli IPMU, KEK, ASIAA, and Princeton University. 

Funding for the Sloan Digital Sky Survey IV has been provided by
the Alfred P. Sloan Foundation, the U.S. Department of Energy Office of
Science, and the Participating Institutions. SDSS-IV acknowledges
support and resources from the Center for High-Performance Computing at
the University of Utah. The SDSS web site is www.sdss.org.

SDSS-IV is managed by the Astrophysical Research Consortium for the 
Participating Institutions of the SDSS Collaboration including the 
Brazilian Participation Group, the Carnegie Institution for Science, 
Carnegie Mellon University, the Chilean Participation Group, the French Participation Group, Harvard-Smithsonian Center for Astrophysics, 
Instituto de Astrof\'isica de Canarias, The Johns Hopkins University, 
Kavli Institute for the Physics and Mathematics of the Universe (IPMU) / 
University of Tokyo, Lawrence Berkeley National Laboratory, 
Leibniz Institut f\"ur Astrophysik Potsdam (AIP),  
Max-Planck-Institut f\"ur Astronomie (MPIA Heidelberg), 
Max-Planck-Institut f\"ur Astrophysik (MPA Garching), 
Max-Planck-Institut f\"ur Extraterrestrische Physik (MPE), 
National Astronomical Observatories of China, New Mexico State University, 
New York University, University of Notre Dame, 
Observat\'ario Nacional / MCTI, The Ohio State University, 
Pennsylvania State University, Shanghai Astronomical Observatory, 
United Kingdom Participation Group,
Universidad Nacional Aut\'onoma de M\'exico, University of Arizona, 
University of Colorado Boulder, University of Oxford, University of Portsmouth, 
University of Utah, University of Virginia, University of Washington, University of Wisconsin, 
Vanderbilt University, and Yale University.

This paper makes use of software developed for the Large Synoptic Survey Telescope. We thank the LSST Project for making their code available as free software at  http://dm.lsst.org.

The Pan-STARRS1 Surveys (PS1) have been made possible through contributions of the Institute for Astronomy, the University of Hawaii, the Pan-STARRS Project Office, the Max-Planck Society and its participating institutes, the Max Planck Institute for Astronomy, Heidelberg and the Max Planck Institute for Extraterrestrial Physics, Garching, The Johns Hopkins University, Durham University, the University of Edinburgh, Queen's University Belfast, the Harvard-Smithsonian Center for Astrophysics, the Las Cumbres Observatory Global Telescope Network Incorporated, the National Central University of Taiwan, the Space Telescope Science Institute, the National Aeronautics and Space Administration under Grant No. NNX08AR22G issued through the Planetary Science Division of the NASA Science Mission Directorate, the National Science Foundation under Grant No. AST-1238877, the University of Maryland, and Eotvos Lorand University (ELTE).
\end{ack}

\end{document}